\documentclass[ reprint,superscriptaddress, amsmath,amssymb, aps,]{revtex4-2}

\usepackage{graphicx}
\usepackage{dcolumn}
\usepackage{bm}
\usepackage[colorlinks=true]{hyperref}
\usepackage{siunitx}
\usepackage{subfigure}
\usepackage[noend]{algorithmic}
\usepackage{algorithm}

\usepackage{xcolor}
\usepackage{soul}
\usepackage{systeme}
\usepackage[capitalize]{cleveref}

\newcommand{\me}{\mathrm{e}}

\newcommand{\red}[1]{\textcolor{red}{#1}}

\usepackage{breakurl}
\begin{document}
\title{Analytical solution to the Poisson-Nernst-Planck equations for the charging of a long electrolyte-filled slit pore}

\author{Timur Aslyamov}
\email{t.aslyamov@skoltech.ru}
\affiliation{Skolkovo Institute of Science and Technology, Bolshoy Boulevard 30, bld. 1, Moscow, Russia 121205}

\author{Mathijs Janssen}
\email{mathijsj@uio.no}
\affiliation{Department of Mathematics, Mechanics Division, University of Oslo, N-0851 Oslo, Norway}
\date{\today}

\begin{abstract}
We study the charging dynamics of a long electrolyte-filled slit pore in response to a suddenly applied potential.
In particular, we analytically solve the Poisson-Nernst-Planck (PNP) equations for a pore for which $\lambda_D\ll H\ll L$, with $\lambda_D$ the Debye length and $H$ and $L$ the pore's width and length.
For small applied potentials, we find the time-dependent potential drop between the pore's surface and its center to be in complete agreement with a prediction of the celebrated transmission line model.
For moderate to high applied potentials, prior numerical work showed that charging slows down at late times;
Our analytical model reproduces and explains such biexponential charge buildup.
\end{abstract}
\maketitle


\section{Introduction}
The behavior of electrolytes in narrow conducting pores and channels is important in various fields of biology, chemistry, as well as in technological applications.
Supercapacitors, for example, store energy through electric double layer (EDL) formation in the nanometer-wide pores of their porous carbon electrodes. 
Such devices are often characterized by measuring the electric current that arises in response to a time-dependent applied potential: be it a step, oscillating (in impedance spectroscopy \cite{lasia2002electrochemical,huang2020review}), or ramps up and down (in cyclic voltammetry \cite{conway2013electrochemical}).
Either way, the microscopic processes that underlie charge storage are measured by these methods only in a volume-averaged manner.

Theoretical models for porous-electrode charging often ignore the complex morphology of these electrodes.
Many molecular simulations, for instance, concern idealized nanometer-sized pore-reservoir system, simulated over nanoseconds \cite{pean2014dynamics,kondrat2014NM,pak2016charging,he2016JPCL,breitsprecher_effect_2017,breitsprecher2018charge,breitsprecher2020speed,bi2020molecular,mo2020ion}. 
As such simulations cannot model the ion transport over millimeters in the quasi-neutral pores of porous electrodes, they vastly underestimate the charging times of real devices \cite{lian2020blessing}.
Other articles numerically solve the Poisson-Nernst-Plank (PNP) equations \cite{sakaguchi2007,lim2009effect,MIRZADEH2014633,PRL2014MTL,henrique2021charging} and dynamical density functional theory (DDFT) \cite{aslyamov2020relation,tomlin2021impedance} to study the charging of cylindrical and slit pores.
As larger length scales could be studied than in MD, the predicted charging times are larger, accordingly.
Yet, a common picture arises from these different numerical methods \cite{sakaguchi2007,pean2014dynamics,kondrat2014NM,breitsprecher2018charge,breitsprecher2020speed, lian2020blessing,aslyamov2020relation}:
Immediately after applying a potential, an electrolyte-filled pore acquires its surface charge diffusively, $\propto \sqrt{t}$, until ionic charge variations penetrate the entire setup and charging goes exponentially with an $RC$ timescale. 
At late times, and especially for large applied potentials, charging slows down and a second exponential regime sets in. 
Before these numerical observations were made, biexponential response had been predicted by Biesheuvel and Bazant's porous electrode model \cite{biesheuvel2010nonlinear}.
As of yet, however, there is no analytical expression based on a comprehensive first-principles derivation that captures biexponential charge build-up.

Decades before porous electrode charging was studied by numerical PNP and molecular simulations, Daniel-Bekh \cite{danielbek1948}, Ksenzhek and Stender \cite{ksenzhek1956}, and de Levie \cite{levie1963,levie1964porous,levie1967electrochemical} developed the transmission line (TL) model. 
The TL model is based on an electronic circuit that distributes the resistance and capacitance of an electrolyte-filled pore over many circuit elements. 
For infinitesimally small circuit elements, the circuit yields a 1d diffusion equation, the TL equation, for the potential drop between the pore's surface and its center \cite{ksenzhek1956,levie1963,janssen2021transmission}.
The response of the TL equation to various potentials and currents was discussed for semi-infinite pores by Ksenzhek and Stender \cite{ksenzhek1956} and de Levie \cite{levie1963} and for finite-length pores in contact with a bulk electrolyte by Posey and Morozumi \cite{posey1966theory}.
The TL impedance found in this way \cite{levie1967electrochemical} has been widely used to fit experimental data \cite{lasia2002electrochemical,huang2020review}.
Likewise, TL model's transient response fitted MD data \cite{bi2020molecular} and accurately reproduced data from numerical solutions of the PNP equations \cite{PRL2014MTL, henrique2021charging}.
Reinforcing the TL model's basis, Henrique, Zuk, and Gupta recently analytically derived the TL equation from the PNP equations \cite{henrique2021charging}.
As they restricted to small applied potentials, however, their model did not capture Biesheuvel and Bazant's late-time slow down.

\begin{figure}
    \centering
    \includegraphics[width=0.45\textwidth]{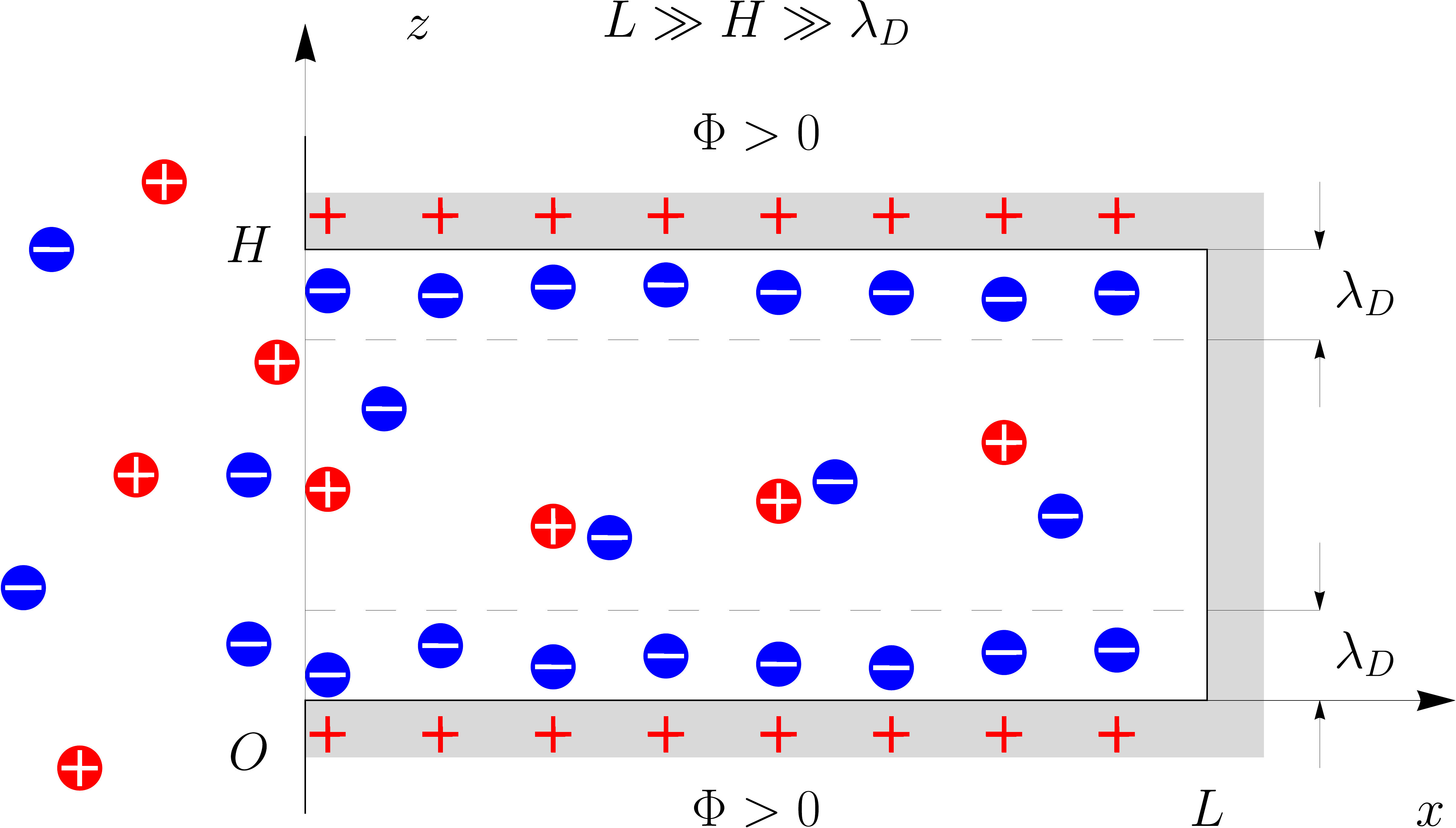}
    \caption{\label{fig:setup} A slit pore subject to an applied potential $\Phi$, closed at the right edge at $x=L$ and in contact at $x=0$ with a bulk filled with a symmetric electrolyte of constant density.
    }
\end{figure}

In this article, we analytically solve the PNP equations to determine the charging dynamics of an electrolyte-filled slit pore (\cref{fig:setup}).
We consider a pore whose length $L$ is greater than its width $H$, which, in turn, is greater than the Debye length $\lambda_D$.
Our derivation hinges on i) asymptotic expansions of the ionic densities and local electrostatic potentials for small $H/L$, which allow us to reduce the 3d PNP equations for the in-pore ion dynamics to a 1d transport equation and ii) an expansion of the time-dependent ionic chemical potentials around the final-state ionic densities.
For small applied potentials, we reproduce Posey and Morozumi's expression for the time-dependent local electrostatic potential inside the pore.
For moderate applied potentials, our model fully explains biexponential surface charge build-up: after initial $RC$-like relaxation, the charging slows down and evolves with the larger diffusion timescale $L^2/D$, with $D$ the ionic diffusion constant.
Our analytically-determined charging times agree with the numerical data of Mirzadeh, Gibou, and Squires \cite{PRL2014MTL}.


\section{Theory}
\subsection{Setup}
We consider the charging of a narrow slit pore with blocking, conducting walls filled with a 1:1 electrolyte. 
The pore's length $L$ is much larger than its width $H$, so that $L\gg H$. 
Moreover, the width is much larger than the size of the ions and solvent molecules and we ignore their finite sizes, accordingly.
We use a Cartesian coordinate system $(x,y,z)$ with $x$ in the length direction and $z$ in the width direction of the pore, see \cref{fig:setup}. 
Moreover, the pore is closed at $x=L$ and in contact with a bulk electrolyte reservoir at salt concentration $c^b$ at $x=0$.
The pore is translationally invariant in the $y$ direction; hence, the dimensionless potential $\phi(t,x,z)$ and the ionic number densities $\rho_{\pm}(t,x,z)$ do not depend on $y$.
From $\phi(t,x,z)$, one finds the local electrostatic potential through multiplication by the thermal voltage $k_B T/e$, with $k_B T$ the thermal energy and $e$ the unit charge.
Likewise, $\rho_{\pm}(t,x,z)$ are the local ionic densities scaled to the bulk ion concentration $c^b$.

We model the evolution of $\rho_\pm(t,x,z)$ and $\phi(t,x,z)$ through the PNP equations, 
\begin{subequations}
\label{eq:general_model}
\begin{align}
    \partial_t \rho_\pm&=D\nabla\cdot \left(\rho_\pm\nabla\mu_\pm\right),\\
    \label{eq:general_model_mu}
    \mu_\pm&=\log(\rho_\pm )\pm \phi, \\
    \nabla^2\phi&=-\frac{\rho_+-\rho_-}{2\lambda_D^2},
\end{align}
\end{subequations}
where $\nabla=(\partial_x, \partial_z)$ is the 2d gradient, where $D$ is the diffusion coefficient, assumed spatially constant and the same for both ion species, where $\lambda_\text{D}=[2c^b e^2/(\varepsilon \varepsilon_0 k_B T)]^{-1/2}$ is the Debye length, with $\varepsilon$ and $\varepsilon_0$ the relative and vacuum permittivity, respectively, and where $\mu_\pm$ are the dimensionless ionic chemical potentials, which are the ionic chemical potentials divided by $k_B T$.

Initially $(t<0)$, no potential is applied to the pore and the electrolyte is homogeneous. 
Charging starts at $t=0$ when the dimensionless surface potential suddenly steps to some nonzero $\Phi$ (not necessarily positive).
\Cref{eq:general_model} is thus subject to the following initial and boundary conditions:
\begin{subequations}
\label{eq:general_boundary_conditions}
\begin{align}
    \label{eq:general_boundary_conditions-ic}
    \rho_\pm(0,x,z)&=1,\\
    \label{eq:general_boundary_conditions-bc}
    \rho_\pm(t,0,z)&=\rho^f_\pm(z),\\
    \phi(t,x,0)&=\Phi.\\
    \phi(t,x,H)&=\Phi,\label{eq:bcupperwall}\\
    \partial_z\mu_\pm(t,x,0)&=0,\label{eq:bcnofluxlower}\\
    \partial_z\mu_\pm(t,x,H)&=0,\label{eq:bcnofluxupper}\\
    \partial_x \mu_\pm(t,L,z)&=0,\label{eq:bcnofluxend}
\end{align}
\end{subequations}
where \cref{eq:bcnofluxlower,eq:bcnofluxupper,eq:bcnofluxend} follow from the pore walls being blocking.
Notice that our setup is symmetric around $z=H/2$.
Hence, from hereon we model only the region $0<z<H/2$ and use $\partial_z\phi(t,x,H/2)=0$ instead of \cref{eq:bcupperwall}.
Notice, also, that we study $\rho_\pm(t,x,z)$ and $\phi(t,x,z)$ only within the pore, $0<x<L$ (and $0<z<H/2$).
In real systems, the potential $\Phi$ is applied with respect to some other electrode.
Especially just after applying the potential, pore charging dynamics can depend on the distance and space between these two electrodes~\cite{janssen2021transmission}\footnote{Personal communication with Jie Yang and Cheng Lian.}.
In our model, however, the reservoir affects the pore only through the boundary condition \cref{eq:general_boundary_conditions-bc} at the orifice ($x=0$).
A key assumption of our model, we postulate that the ionic number densities at $x=0$ relax instantaneously to their final states $\rho^f_\pm(z)$.
As we use the PNP equations, and as we will focus on thin EDLs ($H/\lambda_D\gg1$), these final states are the Gouy-Chapman density profiles 
\begin{equation}\label{eq:GC_density}
    \rho^f_{\pm}(z)=
    \left(\frac{1+\tanh(\Phi/2)\exp(-z/\lambda_D)}{1-\tanh(\Phi/2)\exp(-z/\lambda_D)}\right)^{\mp2}\,.
\end{equation}
The combination of \cref{eq:general_boundary_conditions-bc,eq:GC_density} should be reasonable provided that two conditions are met.
First, the pore should be slender ($L\gg H$), so that slow relaxation in the long in-pore direction allows the system to attain quasi-equilibrium in the short $z$-direction at each time [see \cref{sec:HLexpansion}].
Second, our analysis can only apply to pores whose resistance $R$ is much larger than that of the connected reservoir $R_r$.
For such systems, the electric field drops to zero much faster in the reservoir than in the pore so that the reservoir is in quasi-equilibrium with the pore as it charges. 
Reassuringly, our analysis ultimately reproduces TL results (for the case $R\gg R_r$) \textit{for all times}, implying that the postulated instantaneous densities at $x=0$ are compatible with the TL model.

\subsection{$H/L\ll1$ charging dynamics}\label{sec:HLexpansion}
Instead of fully solving the nonlinear 2d PNP equations \eqref{eq:general_model}, we seek asymptotic approximations to
these equations for small $H/L$, see \cref{sec:appendix_slit-pores}
and Refs.~\cite{aslyamov2020relation,tomlin2021impedance,alizadeh2017multiscale}.
In short, we first rescale the variables in \cref{eq:general_model} with length scales of their characteristic variations.
The scaled PNP equation~\eqref{eq:appendix_general_model} contains $\mathcal{O}(1)$ and $\mathcal{O}(H^2/L^2)$ terms only.
Accordingly, we expand the ionic number densities and dimensionless potential for $H/L\ll 1$ and only retain terms of $\mathcal{O}(1)$ and $\mathcal{O}(H^2/L^2)$, 
\begin{subequations}\label{eq:solution_expansion}
\begin{align}
    \rho_\pm(t,x,z) &= \rho^0_\pm(t,x,z) + \frac{H^2}{L^2} \rho^1_\pm(t,x,z) + \mathcal{O}\left(\frac{H^4}{L^4}\right), \\
    \phi(t,x,z) &= \phi^0(t,x,z) + \frac{H^2}{L^2} \phi^1(t,x,z) + \mathcal{O}\left(\frac{H^4}{L^4}\right).
\end{align}
\end{subequations}
Upon inserting \cref{eq:solution_expansion} into \cref{eq:general_model}, 
we find that the $\mathcal{O}(1)$ problem \cref{eq:appendix_O-1} contains only $z$-derivatives.
In particular, the dimensionless potential $\phi^{0}(t,x,z)$ is governed by 
\begin{subequations}\label{eq:Poisson_equation}
\begin{align}
    \partial_z^2\phi^0&=-\frac{\rho_+^0-\rho_-^0}{2\lambda_D^2},\label{eq:O1Poisson}\\
    \phi^0(t,x,0)&=\Phi,\label{eq:O1Poissonbc1}\\
    \partial_z\phi^0(t,x,H/2)&=0,\label{eq:O1Poissonbc2}
\end{align}
\end{subequations}
Moreover, at $\mathcal{O}(1)$ we find that the chemical potential is constant on $z$-slices of the pore [$\mu_\pm^0(t,x,z)=\mu_\pm^0(t,x)$] throughout the charging process.
The ionic number densities can thus be expressed as [cf.~\cref{eq:general_model_mu}]
\begin{equation}
    \label{eq:density_distribution}
    \rho_\pm^0(t,x,z)=\exp[\mu^0_\pm(t,x)\mp\phi^0(t,x,z)]\,,
\end{equation}
which, inserted into \cref{eq:O1Poisson}, gives 
\begin{align}\label{eq:PNP}
    \partial_z^2 \phi^{0}&=\frac{\exp(\mu^{0}_-+\phi^0)-\exp(\mu^{0}_+-\phi^0)}{2\lambda_D^2}\,.
\end{align}

As the $\mathcal{O}(1)$ problem does not capture the dynamics of our system, we turn to the next order, $\mathcal{O}(H^2/L^2)$, where we find the following transport equation [cf.~\cref{eq:appendix_transport_equation}]: 
\begin{equation}
    \label{eq:transport_average}
    \partial_t \overline{\rho}_\pm^0 - D \partial_x \left(\overline{\rho}_\pm^0 \partial_x \mu_\pm^0\right) = 0,
\end{equation}
where $\overline{\rho}_\pm^0(t,x)$ are cross-sectional averages of the ionic number densities, defined for a general observable $f(t,x,z)$ as
\begin{equation}
    \label{eq:average_density_definition}
    \overline{f}(t,x)=\frac{2}{H}\int_0^{H/2} dz f(t,x,z).
\end{equation}
Notice that, with a slight abuse of notation, we wrote $\overline{\rho}^0(t,x)$ instead of $\overline{\rho^0}(t,x)$, to keep our expressions tractable.

The initial and boundary conditions for \cref{eq:transport_average} follow from cross-sectional averages of \cref{eq:general_boundary_conditions}, 
\begin{subequations}\label{eq:bc_ic}
\begin{align}
    \overline{\rho}^{0}_\pm(0,x)&=1\,,\\
    \overline{\rho}^{0}_\pm(t,0)&=\overline{\rho}^f_\pm\,, \\
    \partial_x \mu_\pm(t,L)&=0\,,
\end{align}
\end{subequations}
where the final-state cross-sectional average densities $\overline{\rho}^f_\pm$ follow from \cref{eq:GC_density,eq:average_density_definition} as
\begin{equation}
    \label{eq:PNP_average_dens_final_1}
    \overline{\rho}_\pm^f=1+ 4 \left[\exp\left(\mp\frac{\Phi}{2}\right)-1\right]\frac{\lambda_D}{H}.
\end{equation}

The key advantages of the $H/L$-expansion are that the transport equation \eqref{eq:transport_average}, which appears at $\mathcal{O}(H^2/L^2)$, is 1d and only contains the first terms of the asymptotic density and potential expansions [\cref{eq:solution_expansion}].
Hence, we do not need to find $\rho^1_\pm(t,x,z)$ and $\phi^1_\pm(t,x,z)$ to characterize the pore's dominant charging dynamics.

This article focuses on analytically solving \cref{eq:transport_average,eq:bc_ic}.
But, for comparison, we also solved these equations numerically, by a procedure outlined below and elaborated upon in \cref{sec:appendix_numerical}.
In our numerical approach, we close \cref{eq:transport_average} by expressing the chemical potential $\mu^0(t,x)\equiv\mu^0[\overline{\rho}^0_+(t,x),\overline{\rho}^0_-(t,x)]$ as functionals of the cross-sectional averages of the ionic number densities $\overline{\rho}^0_\pm(t,x)$.
To do so, we insert \cref{eq:density_distribution} into \cref{eq:average_density_definition} and find
\begin{equation}
    \label{eq:norm}
    \overline{\rho}^0_\pm(t,x)=\overline{\exp\left(\mu^0_\pm\mp\phi^0\right)}=\exp\left(\mu^0_\pm\right)
    \overline{\exp(\mp\phi^0)},
\end{equation}
where, for the second equality, we used that $\mu^0_\pm(t,x)$ is $z$ independent.
With \cref{eq:norm} we rewrite \cref{eq:PNP} to
\begin{equation}
\label{eq:normalized_Poisson}
    \partial_z^2\phi^0=\frac{1}{2\lambda_D^2}\left(\frac{\overline{\rho}^0_-\exp\left(\phi^0\right)}{\overline{\exp\left(\phi^0\right)}}-\frac{\overline{\rho}^0_+\exp\left(-\phi^0\right)}{\overline{\exp\left(-\phi^0\right)}}\right).
\end{equation}

Clearly, a solution $\phi^0(z,\overline{\rho}^0_+, \overline{\rho}^0_-)$ to \cref{eq:normalized_Poisson} is a function of $z$ and of the averaged densities $\overline{\rho}^0_\pm(t,x)$. 
We can thus express the chemical potentials with \cref{eq:norm} as
\begin{equation}
    \label{eq:cp}
    \mu^0_\pm(\overline{\rho}^0_+,\overline{\rho}^0_-)=\log\overline{\rho}_\pm^0-\log\overline{\exp(\mp\phi^0)}\,,
\end{equation}
which depends on the averaged densities $\overline{\rho}^0_\pm(t,x)$ but not on the $z$-coordinate. 
\Cref{eq:cp} enables us to reduce \cref{eq:transport_average} to a closed equation for $\overline{\rho}^{0}_{\pm}(t,x)$.
Details on our numerical implementation are in \cref{sec:appendix_numerical}.

\subsection{Late-time charging dynamics}\label{sec:analytical_model}
We seek an approximate solution to the coupled nonlinear PDE \eqref{eq:transport_average} for times at which the deviations $\delta\overline{\rho}_\pm(t,x)\equiv\overline{\rho}^{0}_\pm(t,x)-\overline{\rho}^f_\pm$ of the densities from their final states are small.
Specifically, we consider Maclaurin series of the density-dependent chemical potentials $\mu^{0}_\pm(\delta\overline{\rho}_+,\delta\overline{\rho}_-)$, omitting terms beyond linear order in $\delta\overline{\rho}_\pm$, we find
\begin{subequations}\label{eq:chemical_potential_expansion_2}
\begin{align}
    \mu^{0}_\pm&=a_\pm \delta\overline{\rho}_+ +b_\pm \delta\overline{\rho}_-+\mathcal{O}(\delta\bar{\rho}_{\pm}^2)
\intertext{ where} 
    \label{eq:aibidefinition}
    a_\pm&=\left.\frac{ \partial\mu^{0}_\pm}{\partial\overline{\rho}_+^0}\right|_{\overline{\rho}^f_+}, \qquad \left. 
    b_\pm=\frac{ \partial\mu^{0}_\pm}{\partial\overline{\rho}_-^0}\right|_{\overline{\rho}^f_-}.
\end{align}
\end{subequations}
Here, we used $\mu^{0}_\pm(\overline{\rho}^f_+, \overline{\rho}^f_-)=0$, which,
for our case of thin EDLs, can be seen from \cref{eq:density_distribution}: at the center of the pore, the potential vanishes, $\phi^0(t,x,H/2)=0$, and final-state density amounts to $\rho^f_{\pm}(z)=1$. 
More general, $\mu^{0}_\pm(\overline{\rho}^f_+, \overline{\rho}^f_-)=0$ follows from the pore being in osmotic contact with a bulk reservoir where $\phi=0$ and $\rho_{\pm}=1$ [cf. \cref{eq:general_model_mu}].

Inserting the linearization \cref{eq:chemical_potential_expansion_2} into \cref{eq:transport_average,eq:bc_ic}, we find 
\begin{subequations}
\label{eq:transport_linear_1}
\begin{align}
    \partial_t\boldsymbol{\delta\overline{\rho}}(t,x)&= D\boldsymbol{A}\partial_x^2\boldsymbol{\delta\overline{\rho}}(t,x)+\mathcal{O}(\delta\bar{\rho}_{\pm}^2), \label{eq:matrix_eq}\\
    \boldsymbol{\delta\overline{\rho}}(0,x)&=\big(1-\overline{\rho}^f_+,1-\overline{\rho}^f_-\big)^{T}, \label{eq:transport_linear_1-ic}\\
    \boldsymbol{\delta\overline{\rho}}(t,0)&=\boldsymbol{0},\\
    \partial_x \boldsymbol{\delta\overline{\rho}}(t,L)&=\boldsymbol{0},
\intertext{where $\boldsymbol{\delta\overline{\rho}}(t,x)=\left(\delta\overline{\rho}_+,\delta\overline{\rho}_-\right)^{T}$ and where}
    \boldsymbol{A}&=
    \begin{pmatrix}
    \overline{\rho}_+^f a_+ & \overline{\rho}_+^f b_+ \\
    \overline{\rho}_-^f a_- & \overline{\rho}_-^f b_-
    \end{pmatrix}. \label{eq:matrix_A_def}
\end{align}
\end{subequations}
As $\overline{\rho}^f_{\pm}$ in \cref{eq:PNP_average_dens_final_1} does not depend on $x$, neither does the initial condition \cref{eq:transport_linear_1-ic}; hence, $\boldsymbol{\delta\overline{\rho}}(0,x)=\boldsymbol{\delta\overline{\rho}}(0)$.

According to the Hartman–Grobman theorem, the behavior of a nonlinear dynamical system of ODEs near a hyperbolic equilibrium point can be described by linearized equations (see Theorem 3.3.1 in Ref.~\cite{arrowsmith1992dynamical}). 
By \cref{eq:transport_linear_1}, we have linearised a nonlinear PDE [\cref{eq:transport_average}], to which that theorem does not apply, but might be extended, see Ref.~\cite{lu1990grobman}. 
Further, our linearization is similar to the linear stability analysis of 1d-DDFT discussed in Section 7.2. of Ref.~\cite{te2020classical} and similar to the chemical-potential expansion of Tomlin and coworkers around a nonhomogenous equilibrium state (Eq. 3.1 in Ref.~\cite{tomlin2021impedance}). 
We have not seen studies of electrolyte dynamics that utilized chemical potential expansions around the final-state densities, though.

As $\mu^0_\pm$ depends only on the cross-sectionally averaged densities, evaluating the derivatives in \cref{eq:aibidefinition} at the $\overline{\rho}^f_{\pm}$ we find that $a_\pm$ and $b_\pm$ are constant determined by the electrolyte properties in the pore at the final state.
Hence, $\boldsymbol{A}$ is constant. 
We assume that matrix $\boldsymbol{A}$ has two distinct real eigenvalues, $\lambda_1$ and $\lambda_2$, and a complete system of eigenvectors, $\boldsymbol{v}_1$ and $\boldsymbol{v}_2$, such that $\boldsymbol{A}\boldsymbol{v}_i=\lambda_i \boldsymbol{v}_i$.
One can thus diagonalize $\boldsymbol{A} = \boldsymbol{P}\boldsymbol{\Lambda} \boldsymbol{P}^{-1}$, where $\boldsymbol{P}=(\boldsymbol{v}_1,\boldsymbol{v}_2)$ and $\boldsymbol{\Lambda} = \mathrm{diag} (\lambda_1, \lambda_2)$, which decouples \cref{eq:transport_linear_1} to 
\begin{equation}
    \label{eq:appendix_transport_linear_2}
    \partial_t g_i(t,x)= \lambda_i D \partial_x^2g_i(t,x),
\end{equation}
where $g_i$ are the components of the vector $\boldsymbol{g}=\boldsymbol{P}^{-1}\boldsymbol{\delta\overline{\rho}}$. 
Notice that, to write \cref{eq:appendix_transport_linear_2}, we have used that $\boldsymbol{A}$ does not depend on time. 
The following boundary and initial conditions apply:
\begin{subequations}\label{eq:appendix_bc_ic_linear}
\begin{align}
    g_i(0)&=\left(\boldsymbol{P}^{-1}\boldsymbol{\delta\overline{\rho}}(0)\right)_{i},\label{eq:gi_initial}\\
    g_i(t,0)&=0, \\
    \partial_x g_i(t,L)&=0.
\end{align}
\end{subequations}
Notice that the initial condition \eqref{eq:gi_initial} does not depend on $x$, as neither $\boldsymbol{P}^{-1}$ nor $\boldsymbol{\delta\overline{\rho}}(0)$ does.
\Cref{eq:appendix_transport_linear_2,eq:appendix_bc_ic_linear} represent a standard heat conduction problem that can be solved with separation of variables \cite{whitaker2013fundamental}, Laplace transformations, or Green's functions. 
We found
\begin{equation}
    \label{eq:appendix_transport_solve_2}
    g_i(t,x)=\sum_{n=0}^{\infty}\frac{2\sin(\beta_n x/L)}{\beta_n} \exp(-\beta_n^2 \lambda_i Dt/L^2)g_i(0)\,,
\end{equation}
where $\beta_n=\pi\left(1/2+ n\right)$.
In vector form, \cref{eq:appendix_transport_solve_2} reads
\begin{align}
    \label{eq:appendix_transport_solve_3}
    &\boldsymbol{g}(t,x)=\sum_{n=0}^\infty\frac{2 \sin(\beta_n x/L)}{\beta_n}\nonumber\\
    &\quad \mathrm{diag}\!\left[\exp\left(-\beta_n^2\lambda_1 \frac{Dt}{L^2}\right),\exp{\left(-\beta_n^2\lambda_2 \frac{Dt}{L^2}\right)}\right]\boldsymbol{g}(0)\,.
\end{align}
We calculate the density variation $\boldsymbol{\delta\overline{\rho}}=\boldsymbol{P}\boldsymbol{g}$ and, with the matrix exponent identity, 
\begin{equation}
    \label{eq:appendix_matrix_exponent_identity}
    \exp{\left(\boldsymbol{A}\right)}=\boldsymbol{P}\,\mathrm{diag}\bm{(}\exp{(\lambda_1) },\exp{(\lambda_2)}\bm{)}\,\boldsymbol{P}^{-1},
\end{equation}
we find the following solution to \cref{eq:transport_linear_1}:
\begin{equation}
    \label{eq:density_variance_vector}
    \boldsymbol{\delta\overline{\rho}}(t,x)=\sum_{n=0}^\infty\frac{2 \sin(\beta_n x/L)}{\beta_n}\exp{\left(-\beta_n^2 \boldsymbol{A}\frac{Dt}{L^2} \right)}\boldsymbol{\delta\overline{\rho}}(0).
\end{equation}

A key quantity capturing the charging state of a pore is its length-averaged charge density, $Q(t)=\int_0^L dx(\overline{\rho}^{0}_+-\overline{\rho}^{0}_-)/L$.
For our setup with like-charged pore walls, $Q(t)$ is opposite and equal to the wall-averaged electric surface charge density.
Instead of on $Q(t)$, we will focus on the deviation $\delta Q(t)=Q^f-Q(t)$ from its final value $Q_f=\overline{\rho}_+^f-\overline{\rho}_-^f$. 
In terms of $\boldsymbol{\delta\overline{\rho}}$ and the ionic valency vector $\boldsymbol{z}=\left(1,-1\right)^{T}$ we find
\begin{equation}
    \label{eq:charge_dynamics_def}
    \delta Q(t)=-\frac{1}{L}\int_0^L dx \boldsymbol{z}^{T}\boldsymbol{\delta \overline{\rho}}(t,x).
\end{equation}
Inserting \cref{eq:density_variance_vector} into \cref{eq:charge_dynamics_def} then yields
\begin{equation}
    \label{eq:transport_solve_5_full}
    \delta Q(t)=-\sum_{n=0}^\infty\frac{2}{\beta_n^2} \sum_{i=1}^2\boldsymbol{z}^T\boldsymbol{A}_i\boldsymbol{\delta\overline{\rho}}(0)\exp{\left(-\beta_n^{2}\lambda_i \frac{ Dt}{L^2}\right)},
\end{equation}
where we used Sylvester's formula, 
\begin{equation}
    \label{eq:Sylvester}
    \exp{\left(-\beta_n^{2}\boldsymbol{A}\frac{Dt}{L^2} \right)}=\sum_{i=1}^2\boldsymbol{A}_i\exp{\left(-\beta_n^{2}\lambda_i\frac{ Dt}{L^2} \right)},
\end{equation}
and the Frobenius covariants $\boldsymbol{A}_i$, 
\begin{equation}
    \label{eq:Frobenius}
    \boldsymbol{A}_1=\frac{\boldsymbol{A}-\lambda_1\boldsymbol{I}}{\lambda_1-\lambda_2},\qquad
    \boldsymbol{A}_2=\frac{\boldsymbol{A}-\lambda_2\boldsymbol{I}}{\lambda_2-\lambda_1},
\end{equation}
with $\boldsymbol{I}$ the identity matrix. 

Except at early times, $\delta Q(t)$ is dominated by its $n=0$ terms, which relax with timescales $\propto L^2/(D\lambda_1)$ and \mbox{$\propto L^2/(D\lambda_2)$}.
Hence, both timescales go as $L^2/D$, which corresponds to electrolyte diffusion along the length of the pore.
But these timescales can still differ much through the factors $\lambda_1$ and $\lambda_2$, which depend on the pore and electrolyte properties through the coefficients $a_\pm$ and $b_\pm$ [\cref{eq:aibidefinition}]. 

\subsection{Analytical approximations to $\delta Q(t)$ for thin EDLs and moderate potentials}\label{sec:analytical_model_thin_layer}
We will seek analytical expressions for $\delta Q(t)$ [\cref{eq:transport_solve_5_full}] by considering increasingly-restrictive constraints on the values of $\Phi$ and $\lambda_D/H$. 

\subsubsection{Thin double layers: $\exp(-H/\lambda_D)\ll 1$}
We seek a solution to \cref{eq:PNP} and start by splitting $\phi^{0}(t,x,z)$ into
\begin{equation}
    \label{eq:PNP_solution_1}
    \phi^{0}(t,x,z)=\frac{\mu_+^{0}(t,x)-\mu_-^{0}(t,x)}{2}+\phi^{\rm GC}(t,x,z)\,,
\end{equation}
where the superscript tentatively refers to Gouy and Chapman.
Inserting this expression into \cref{eq:PNP}, we find that $\phi^{\rm GC}(t, x, z)$ is governed by
\begin{subequations}\label{eq:GC_equation}
\begin{align}
    \partial_z^2 \phi^{\rm GC}&=\lambda_m^{-2}
    \sinh\left(\phi^{\rm GC}\right),\\
    \phi^{\rm GC}(t,x,0)&=\Phi_m,\\
    \partial_z\phi^{\rm GC}(t,x, H/2)&=0,
\end{align}
\end{subequations}
where $\Phi_m=\Phi-(\mu_+^{0}-\mu_-^{0})/2$ is a modified dimensionless surface potential and $\lambda_m=\lambda_D\exp[-(\mu_+^{0}+\mu_-^{0})/4]$ a modified Debye length, which both depend on $t$ and $x$ through $\mu_{\pm}^{0}$. 
For general $H/\lambda_m$, an equation equivalent to \cref{eq:GC_equation} was solved by Corkill and Rosenhead \cite{corkill1939distribution}, with a solution [Eq.~(3.7) therein] in terms of elliptical functions. 
Meanwhile, for pores much wider than the Debye length, the Poisson-Boltzmann equation \eqref{eq:GC_equation} has the famous Gouy-Chapman solution 
\begin{equation}
    \label{eq:PNP_solution_GC}
    \phi^{\rm GC}(t,x,z)=2\log\frac{1+\tanh\left(\Phi_m/4\right)\exp(- z/\lambda_m)}{1-\tanh\left(\Phi_m/4\right)\exp(-z/\lambda_m)}+\dots\,.
\end{equation}
Here, dots represent higher order terms in an $H/\lambda_D\ll1$ expansion of the elliptic functions of Corkill and Rosenhead \cite{corkill1939distribution}. 
They showed that such terms are negligible for $H/\lambda_D>16$; the smallest value of that fraction that we consider here is $H/\lambda_D=20$. 
Inserting \cref{eq:PNP_solution_GC} into \cref{eq:PNP_solution_1} then yields the solution of \cref{eq:PNP}.
As $\phi^{\rm GC}(t, x, H/2)=0$, we see from \cref{eq:PNP_solution_1} that non-equal chemical potentials $\mu_+^{0}\neq\mu_-^{0}$ result in a nonzero potential at the middle of the pore.

Next, we insert \cref{eq:PNP_solution_1,eq:PNP_solution_GC,eq:density_distribution} into \cref{eq:average_density_definition} to determine the cross-sectional average densities,
\begin{align}
    \label{eq:PNP_average_dens_2}
    \overline{\rho}^{0}_\pm&=\exp\left(\frac{\mu_+^0+\mu_-^0}{2}\right)\left\{1 +\frac{4\lambda_m}{H}\left[ \exp\left(\mp\frac{\Phi_m}{2}\right)-1 \right]\right\},\nonumber\\
    &\qquad+\mathcal{O}\left(\exp(-H/\lambda_m)\right)\,,
\end{align}
where the neglected higher order terms stem from the leading order term of \cref{eq:PNP_solution_GC}.
As $\lambda_m$ scales as $\lambda_D$ at the linear expansion near the final state, $\lambda_m=\lambda_D[1+\mathcal{O}(\delta\overline{\rho}_\pm)]$, we find that neglecting these terms in \cref{eq:PNP_average_dens_2} means that our theory holds for moderately thin ELDs [$O\bm{(}\exp(-H/\lambda_D)\bm{)}$].
(Notice that the final-state cross-sectional average densities $\overline{\rho}^f_\pm$ of \cref{eq:PNP_average_dens_final_1} also follow from setting $\mu_\pm=0$, $\lambda_m=\lambda_D$ and $\Phi_m=\Phi$ in \cref{eq:PNP_average_dens_2}.)

Writing $\overline{\rho}^{0}_\pm=\overline{\rho}^{0}_\pm(\mu^{0}_+,\mu^{0}_-)$ and differentiating both sides with respect to $\overline{\rho}_\pm^{0}$, we obtain four independent equations for $a_\pm$ and $b_\pm$,
\begin{subequations}
\label{eq:appendix_aibi_equations}
\begin{align}
1&=\frac{\partial \overline{\rho}^0_+}{\partial\mu^0_+}a_++\frac{\partial \overline{\rho}^0_+}{\partial\mu^0_-}a_-,\\ 
0&=\frac{\partial \overline{\rho}^0_+}{\partial\mu^0_+}b_++\frac{\partial \overline{\rho}^0_+}{\partial\mu^0_-}b_-,\\
1&=\frac{\partial \overline{\rho}_-}{\partial\mu_+}b_++\frac{\partial \overline{\rho}_-}{\partial\mu_-}b_-,\\
0&=\frac{\partial \overline{\rho}_-}{\partial\mu_+}a_++\frac{\partial \overline{\rho}_-}{\partial\mu_-}a_-.
\end{align}
\end{subequations}
Inserting \cref{eq:PNP_average_dens_2}, we find 
\begin{subequations}\label{eq:aibi_result}
\begin{align}
    a_-&=-\frac{(H-2\lambda_D)H}{4\lambda_D\left[2\lambda_D+(H-2\lambda_D) \cosh(\Phi/2)\right]}, \\
    a_+&=-a_-\left(1+\frac{4\lambda_D\exp\left(\Phi/2\right)}{H-2\lambda_D}\right),\\
    b_-&=-a_-\left(1+\frac{4\lambda_D\exp\left(-\Phi/2\right)}{H-2\lambda_D}\right),\\
    b_+&=a_-\,.
\end{align}
\end{subequations}
With the coefficient $a_\pm$ and $b_\pm$ [\cref{eq:appendix_aibi_equations}] and final densities $\rho_{\pm}^f$ [\cref{eq:PNP_average_dens_final_1}] at hand, we can now express $\boldsymbol{A}$ [\cref{eq:matrix_A_def}] and analytically determine its eigenvalues and Frobenius covariants $\boldsymbol{A}_i$ [\cref{eq:Frobenius}].
In turn, this yields the charging dynamics $\delta Q(t)$ [\cref{eq:transport_solve_5_full}]. 
The resulting expressions, however, are very long (not shown).

\subsubsection{Thin double layers and moderate potentials: $\lambda_D/H\ll1$, $\exp{(\pm\Phi/2)}\lambda_D^2/H^2\ll1$}
We further restrict the EDL thickness and also constrain the applied potential by omitting $\mathcal{O}(\lambda_D/H)$ and $\mathcal{O}\bm{(}\exp{(\pm\Phi/2)}\lambda_D^2/H^2\bm{)}$ terms.
We do keep $\mathcal{O}\bm{(}\exp{(\pm\Phi/2)}\lambda_D/H\bm{)}$ terms, which can become notable for $|\Phi|>1$.
Under these conditions, \cref{eq:aibi_result} reduces to
\begin{subequations}
\label{eq:aibi_PNP_1}
\begin{align}
    a_-&=-\frac{1}{4\cosh(\Phi/2)}\frac{H}{\lambda_D}\left[1+\mathcal{O}\left(\frac{\lambda_D}{H}\right)\right]\,,  \label{eq:aibi_PNP_1a} \\
    a_+&=-a_-\left\{1+\eta\left[1+\mathcal{O}\left(\frac{\lambda_D^2}{H^2}\right)\right]\right\}\,,                 \label{eq:aibi_PNP_1b}\\
    b_-&=-a_-\left\{1+\exp(-\Phi)\eta\left[1+\mathcal{O}\left(\frac{\lambda_D^2}{H^2}\right)\right]\right\}\,,      \label{eq:aibi_PNP_1c}\\
    b_+&=a_-
\end{align}
\end{subequations}
where the parameter $\eta$ is given by 
\begin{equation}
    \label{eq:dukhin_number}
    \eta=4\exp\left(\frac{\Phi}{2}\right)\frac{\lambda_D}{H}.
\end{equation}
Parameters similar to $\eta$ appear in models for electrophoresis (as the ``Dukhin" number) \cite{dukhin1993non} and EDL formation near flat plates \cite{bazant2004diffuse,kilic2007steric}.

Using \cref{eq:PNP_average_dens_final_1} and omitting $\mathcal{O}(\lambda_D/H)$ and $\mathcal{O}\bm{(}\exp{(\pm\Phi/2)}\lambda_D^2/H^2\bm{)}$ terms, we find $\overline{\rho}_+^f a_+=\overline{\rho}_-^f b_-$, with $a_+$ and $b_-$ given in \cref{eq:aibi_PNP_1}. 
Inserting this into \cref{eq:matrix_A_def} yields 
\begin{equation}
    \label{eq:PNP_matrix_A}
    \boldsymbol{A}=\frac{1}{4 \cosh(\Phi/2)}\frac{H}{\lambda_D}
    \begin{pmatrix}
    \overline{\rho}_+^f(1+\eta) & -\overline{\rho}_+^f \\
    -\overline{\rho}_-^f & \overline{\rho}_+^f(1+\eta)
    \end{pmatrix},
\end{equation}
whose eigenvalues and eigenvectors read
\begin{subequations}
\begin{align}
    \lambda_1&=\frac{1}{4 \cosh(\Phi/2)}\frac{H}{\lambda_D}\left[\overline{\rho}_+^f(1+\eta)+\sqrt{1+\overline{\rho}_\text{exc}}\right]\,,\label{eq:PNP_ev_1}\\
    \lambda_2&=\frac{1}{4 \cosh(\Phi/2)}\frac{H}{\lambda_D}\left[\overline{\rho}_+^f(1+\eta)-\sqrt{1+\overline{\rho}_\text{exc}}\right]\,,\label{eq:PNP_ev_2}\\
    \boldsymbol{v}_1&=\left(-\sqrt{\overline{\rho}_+^f/\overline{\rho}_-^f}, 1\right)^{T},                     \label{eq:PNP_evectors}\\
    \boldsymbol{v}_2&=\left(\sqrt{\overline{\rho}_+^f/\overline{\rho}_-^f}, 1\right)^{T},
\end{align}
\end{subequations}
where $\overline{\rho}_\text{exc}=\overline{\rho}_+^f+\overline{\rho}_-^f-2$ is the excess salt density, which arises in the eigenvalues \cref{eq:PNP_ev_1,eq:PNP_ev_2} as $\overline{\rho}_+^f\overline{\rho}_-^f=1+\overline{\rho}_\text{exc}+\mathcal{O}\bm{(}\exp{(\pm\Phi/2)}\lambda_D^2/H^2\bm{)}$. 

Calculating the Frobenius covariants [\cref{eq:Frobenius}] and inserting them into \cref{eq:transport_solve_5_full}, we obtain
\begin{subequations}
\label{eq:analytic_charging_charge-salt}
\begin{align}
    \label{eq:analytic_charging_full}
    \frac{\delta Q(t)}{Q^f}&=\sum_{n=0}^\infty\frac{1}{\beta_n^2}
    \left\{
    \left[1+(1+\overline{\rho}_\text{exc})^{-1/2}\right]\exp{\left(-\beta_n^{2}\lambda_1\frac{Dt}{L^2}\right)}\right.\nonumber\\
    &\qquad+\left.\left[1-(1+\overline{\rho}_\text{exc})^{-1/2}\right]\exp{\left(-\beta_n^{2}\lambda_2\frac{Dt}{L^2}\right)}
    \right\}\nonumber\\
    &\qquad+\mathcal{O}\left(\frac{\lambda_D}{H}\right) +\mathcal{O}\left(\me^{\pm\Phi}\lambda_D^2/H^2\right),
\intertext{where}
    \overline{\rho}_\text{exc}&=16\frac{\lambda_D}{H}\sinh^2(\Phi/4)\label{eq:salt_exc}
\end{align}
\end{subequations}
Notice that the relative importance of the two terms of \cref{eq:analytic_charging_full} depends only on $\overline{\rho}_\text{exc}$, which, in turn, depends on the applied potential and EDL overlap. 

\subsubsection{Towards the TL model: $\eta\ll1$}\label{sec:towarrdsTL}
Next, we consider the case $\eta\ll 1$.
Clearly, for $\eta$ to be a small parameter puts restrictions on the applied potential $\Phi$ and the EDL overlap $\lambda_D/H$.
Yet, $\eta\ll 1$ and $\Phi\sim1$ are simultaneously possible.
Thus, for sufficiently thin EDLs, the expressions that we derive below apply to PNP in the nonlinear charging regime.

We insert \cref{eq:PNP_average_dens_final_1,eq:salt_exc} for $\overline{\rho}_\pm^f$ and $\overline{\rho}_\text{exc}$ into \cref{eq:PNP_ev_1} to obtain small-$\eta$ expansions of the eigenvalues,
\begin{subequations}\label{eq:PNP_evalues_linear}
\begin{align}
    \lambda_1&=\frac{2}{\eta}\frac{\exp(\Phi/2)}{\cosh(\Phi/2)}+\mathcal{O}(1),\\
    \lambda_2&=1+\mathcal{O}(\eta).
\end{align}
\end{subequations}
Likewise, we find that \cref{eq:analytic_charging_charge-salt} reduces for small $\eta$ to
\begin{align}\label{eq:PNP_charge_dynamics}
    \frac{\delta Q}{Q^f}&=\sum_{n=0}^\infty\frac{2}{\beta_n^2}\left[\left(1-\frac{\sinh^2(\Phi/4)}{\exp(\Phi/2)}\eta \right)\exp{\left(-\beta_n^{2}\lambda_1\frac{Dt}{L^2}\right)}\right.\nonumber\\
    &\qquad\qquad+\left.\eta\frac{\sinh^2(\Phi/4)}{\exp(\Phi/2)}\exp{\left(-\beta_n^{2}\lambda_2\frac{Dt}{L^2}\right)}\right] +\mathcal{O}(\eta^2)
\end{align}
At late times, only the $n=0$ terms contribute and $\delta Q(t)$ further simplifies to
\begin{subequations}\label{eq:PNP_charge_dynamics_late_time}
\begin{align}
    \frac{\delta Q}{Q^f}&\simeq\frac{8}{\pi^2}\left[\exp{\left(-\frac{t}{\tau_1}\right)}+\eta\frac{\sinh^2(\Phi/4)}{\exp(\Phi/2)}\exp{\left(-\frac{t}{\tau_2}\right)}\right] \label{eq:PNP_charge_dynamics1}\\
    \tau_1&=\frac{4}{\pi^2}\frac{L^2}{D} \frac{2\lambda_D}{H}\cosh\left(\frac{\Phi}{2}\right), \label{eq:PNP_charging_time_1}\\
    \tau_2&=\frac{4}{\pi^2}\frac{L^2}{D}\gg\tau_1\,,\label{eq:PNP_charging_time_2}
\end{align}
\end{subequations}
where we omitted a $\mathcal{O}(\eta)$ term in the term relaxing with $\tau_1$ relaxation as it is much smaller than the $\mathcal{O}(1)$ term that we kept. 
We kept the $\mathcal{O}(\eta)$ term that relaxes with the $\tau_2$ timescale, however, as it can dominate the first term of \cref{eq:PNP_charge_dynamics1} for $\tau_1<t<\tau_{2}$.
In $\tau_2$ we recognise the common diffusion timescale; as we considered $\lambda_D\ll H$, it follows that $\tau_2\gg\tau_1$. 
Next, we understand $\tau_1$ as follows.
Multiplying the differential Gouy-Chapman capacity per unit length in the $y$-direction, $C_\text{D}^\text{GC}=2L\varepsilon\varepsilon_0\cosh(\Phi/2)/\lambda_D$, by the electrolyte resistance times a unit length in the $y$-direction, $R=\lambda_D^2L/(H\varepsilon\varepsilon_0D)$, yields the timescale $RC=2\lambda_DL^2/(HD)\cosh(\Phi/2)$.
To the best of our knowledge, this timescale has not been reported for pores.
Yet, it is completely analogous to the nonlinear \textit{RC} time of flat-electrode charging \cite{bazant2004diffuse}.
In both cases, the nonlinear $RC$ time comprises a $\Phi$-independent prefactor multiplied by $\cosh(\Phi/2)$.

\Cref{eq:PNP_charge_dynamics} and its late-time simplification \cref{eq:PNP_charge_dynamics_late_time} are key results of this paper.
As we will discuss further in \cref{sec:results,sec:discussion}, these analytical expressions fully capture the biexponential charge relaxation seen in previous numerical works \cite{kondrat2014NM,breitsprecher2018charge,breitsprecher2020speed,pean2014dynamics}.

\subsubsection{TL model: $\Phi\ll1$}\label{sec:TL}
For $\Phi\ll1$ our theory recovers known TL model results.
First, the timescale $\tau_1$ reduces to
\begin{equation}
    \label{eq:PNP_charging_time_TL}
    \tau_1=\frac{4}{\pi^2}\frac{L^2}{D}\frac{\lambda_D}{h_p},
\end{equation}
where $h_p\simeq H/2$ is the ratio of the pore's cross-section area to perimeter for narrow pores $H/L\ll 1$.
Apart from the prefactor $4/\pi^2$, the above timescale agrees with the timescale $\tau_{TL}=L^2\lambda_D/(h_p D)$ of Ref.~\cite{PRL2014MTL}. 

Second, dropping the $\mathcal{O}(\eta)$ terms in \cref{eq:PNP_charge_dynamics} yields
\begin{equation}
    \label{eq:charge_dynamics_TL}
    \frac{\delta Q(t)}{Q^f}=\sum_{n=0}^\infty\frac{2}{\beta_n^2}\exp\left(-\frac{\beta_n^2 t}{\tau_{TL}}\right),
\end{equation}
which is the charge density stated below Eq.~(7) in Ref.~\cite{PRL2014MTL}.

Last, we consider the electrostatic potential difference between the pore’s surface and center line, $\Delta\phi(t,x)=\Phi-\phi^0(t,x,H/2)$.
As the Gouy-Chapman potential $\phi^{\rm GC}$ [\cref{eq:PNP_solution_GC}] vanishes at the mid-plane ($z=H/2$), with \cref{eq:PNP_solution_1} we find that $\Delta\phi=\Phi-(\mu^0_+-\mu^0_-)/2$.
Using \cref{eq:aibi_PNP_1,eq:chemical_potential_expansion_2} and $a_-=-H/(4\lambda_D)$ for $\Phi\ll1$, we find
\begin{align}
    \label{eq:potential_difference_1}
    \frac{\Delta\phi(t,x)}{\Phi}&=1-\frac{H\boldsymbol{z}^T\boldsymbol{\delta\overline{\rho}}(t,x)}{4\lambda_D\Phi},
\end{align}
We use \cref{eq:density_variance_vector,eq:Sylvester} to determine $\boldsymbol{z}^T\boldsymbol{\delta\overline{\rho}}(t,x)$. 
In this calculation, the first Frobenius covariant contributes with a term $\boldsymbol{z}^T\boldsymbol{A}_1\boldsymbol{\delta\overline{\rho}}(0)=4\Phi \lambda_D/H+\mathcal{O}(\eta^2)$, while the second covariant $\boldsymbol{A}_2\sim\mathcal{O}(\eta^2)$ is discarded. 
We find
\begin{align}
    \label{eq:TLpotential_difference}
    \frac{\Delta\phi(t,x)}{\Phi}&=1-\sum_{n=0}^\infty\frac{2\sin(\beta_n x/L)}{\beta_n}\exp\left(-\frac{\beta_n^2 t}{\tau_{TL}}\right),
\end{align}
which coincides with Eq.~(19) of Ref.~\cite{posey1966theory}.
Underlying our derivation of \cref{eq:TLpotential_difference} is the assumption that the ion densities at $x=0$ relaxed instantaneously [\cref{eq:general_boundary_conditions-bc}] to the Gouy-Chapman density [\cref{eq:GC_density}]. 
Notice that $\Delta \phi(t=0,x=0)=\Phi$ for these Gouy-Chapman densities, which was precisely the boundary condition used by Ref.~\cite{posey1966theory} to derive their Eq.~(19).

To our knowledge, we have thus given the first comprehensive derivation of TL-model results starting from first principles.


\section{Results}\label{sec:results}
\begin{figure}
    \centering
    \includegraphics[width=0.47\textwidth]{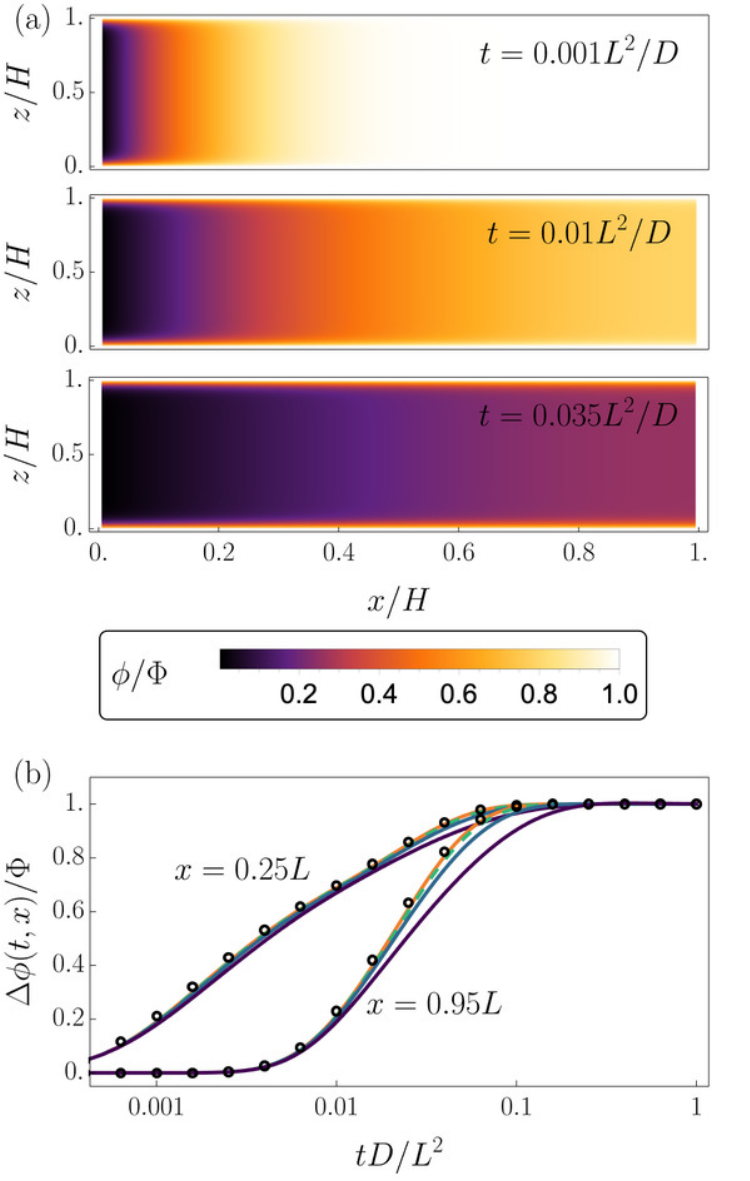}
    \caption{
    (a) Heat map of $\phi^0(t,x,z)$ for $\Phi=1$ and three different times; for this plot, we used \cref{eq:PNP_solution_1,eq:PNP_solution_GC} and determined $\mu^0_{\pm}$ by a semi-analytical described in the text.
    (b) The potential drop between the pore surface and pore center, $\Delta\phi(x,t)=\Phi-\phi^0(t,x,H/2)$, as a function of $t$ at $x=0.25L$ and $x=0.95L$, for $H/\lambda_D=40$ and $\Phi=0.1, 1.0, 2.0$ and $4.0$ (orange, green dashed, blue, and purple lines).
    Lines are determined by the semi-analytical method; black open circles are the TL solution \cref{eq:potential_difference_1}. 
    }
    \label{fig:TLM_potential}
\end{figure}

We first discuss the dimensionless potential $\phi^0(t,x,z)$ [\cref{eq:PNP_solution_1}] inside our slit pore.
To plot that equation requires inserting $\phi^{\rm GC}$ [\cref{eq:PNP_solution_GC}] and $\mu^0_{\pm}(\overline{\rho}^0_+,\overline{\rho}^0_-)$---the latter quantity we determined by a semi-analytical method whereby we evaluated \cref{eq:chemical_potential_expansion_2} with numerically-determined coefficients $a_\pm$ and $b_\pm$, see \cref{sec:appendix_numerical}.
\Cref{fig:TLM_potential}(a) shows heat maps of $\phi^0(t,x,z)$ from \cref{eq:PNP_solution_1} as it evolves inside the slit pore. 
In these snapshots, we see how a ``charging front" penetrates the pore.
Our semi-analytical model thus contains more information than the TL model, which only describes the dynamics of the electrostatic potential drop $\Delta \phi(t,x)=\Phi-\phi^0(t,x,H/2)$ between the the pore surface and its mid-plane, which we turn to next.
\Cref{fig:TLM_potential}(b) shows $\Delta \phi(t,x)/\Phi$ as determined by the semi-analytical method for $x=0.25L$ and $x=0.95L$, $H/\lambda_D=40$, and several $\Phi$.
The same panel also shows the $\Phi$-independent TL solution \cref{eq:TLpotential_difference} (black open circles), 
in whose derivation we omitted $\mathcal{O}(\eta^2)$ terms of $\boldsymbol{A}_1$ and $\boldsymbol{A}_2$, which would have contributed to $\Delta \phi(t,x)/\Phi$ at $\mathcal{O}(\eta)$.
The plot shows that \cref{eq:TLpotential_difference} agrees with the semi-analytical results up to $\Phi=1$, for which $\eta\approx0.16$ is indeed small. 
This agreement up to $\Phi=1$ is surprising on the basis the TL equation's usual derivation, which involves a $\Phi\ll 1$ assumption \cite{PRL2014MTL}.

\begin{figure}
\includegraphics[width=0.47\textwidth]{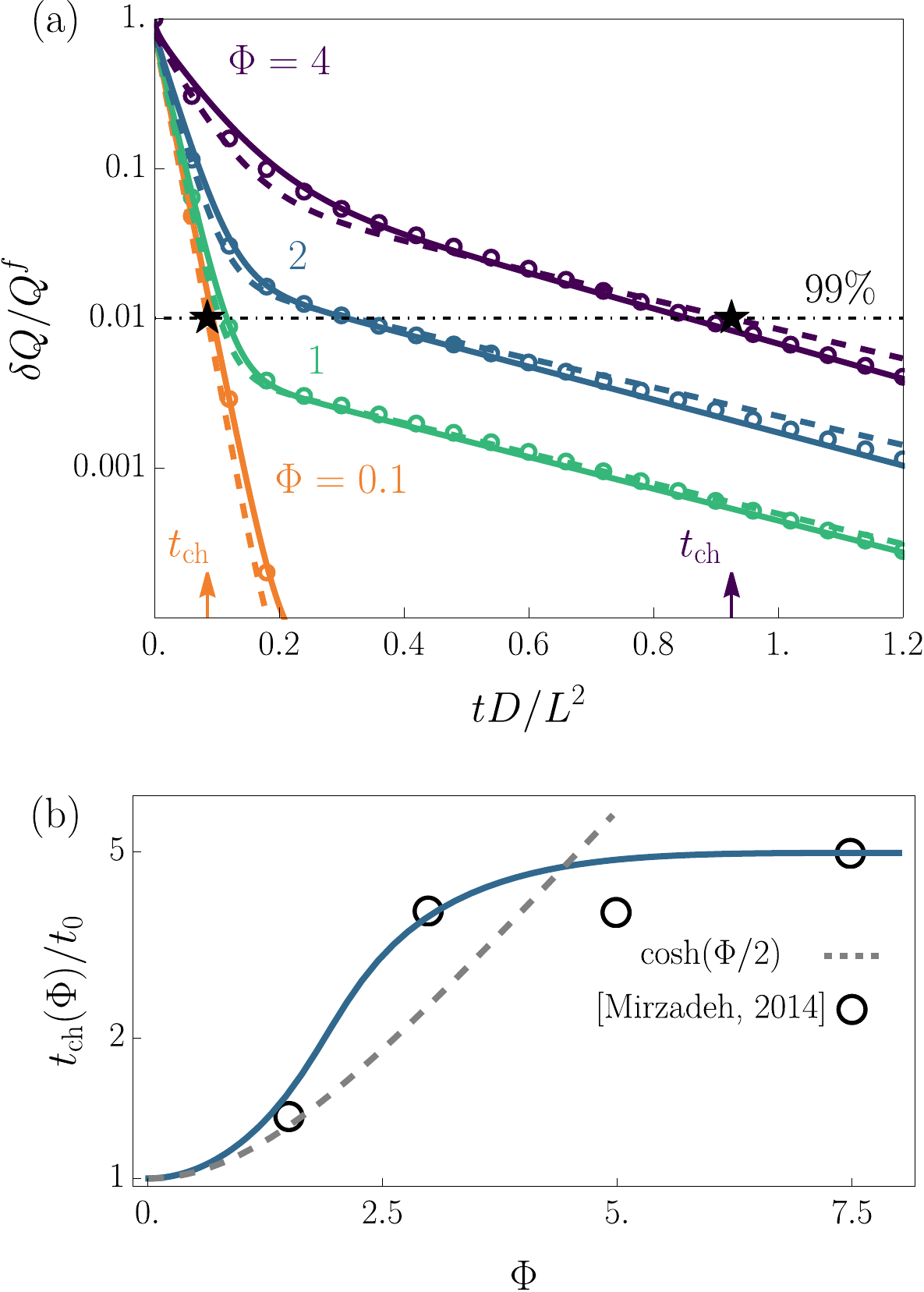}
    \caption{
    (a) Charging of a slit pore with $H/\lambda_D=40$ after applying a potential $\Phi=0.1, 1.0, 2.0$ or $4.0$ (orange, green, blue, purple), calculated from the transport equations \eqref{eq:transport_average} and \eqref{eq:bc_ic} numerically (open circles), semi-analytically (solid lines) and analytically [\cref{eq:analytic_charging_charge-salt}] (dashed lines). 
    The dot-dashed line illustrates point where the system reaches $99\%$ of its charge. 
    The vertical arrows indicate, for $\Phi=0.1$ and $\Phi=4.0$, the times $t_\text{ch}$ at which that barrier is crossed.
    (b) The scaled charging time ratio $t_\text{ch}/t_0$ (blue line) for $H/\lambda_D=20$, calculated semi-analytically using \cref{eq:transport_solve_5_full}. 
    The black open circles are data from Fig.~5(d) of Ref.~\cite{PRL2014MTL}, obtained there through numerical solution of the PNP equations. 
    The gray dashed line indicates $\cosh(\Phi/2)$.
    }
    \label{fig:charging_dynamics}
\end{figure}

Next, we discuss the deviation of the average ionic charge density from its final state, $\delta Q(t)$.
The theory of the previous section enables us to determine $\delta Q(t)$ at different levels of restrictions to the parameters $\lambda_D/H$ and $\Phi$.
Here, we choose the following three methods to determine $\delta Q(t)$:
i) numerically, by solving \cref{eq:transport_average,eq:bc_ic}, see \cref{sec:appendix_numerical}; 
ii) semi-analytically (in the same way as we determined $\phi^0(t,x,H/2)$ above), with \cref{eq:transport_solve_5_full} and numerically determined $a_{\pm}$ and $b_{\pm}$, see \cref{sec:appendix_numerical};
iii) analytically, with \cref{eq:analytic_charging_charge-salt}.
Accordingly, \cref{fig:charging_dynamics}(a) shows numerical (open circles), semi-analytical (lines), and analytical (dashed lines) results for $\delta Q(t)$ for several $\Phi$. 
Comparing the results of the three methods, we see that the numerical and semi-analytical methods yield almost indistinguishable $\delta Q(t)$; predictions from \cref{eq:analytic_charging_charge-salt} differ a bit, but still agree with the other methods within a few percent.
Clearly, all three methods predict the same qualitative behavior:
For the small value $\Phi=0.1$, the charge evolves with a single characteristic time; for $\Phi\ge1$, the charge relaxes exponentially with two distinct timescales.
For all $\Phi$ considered, the first exponential regime describes almost the whole charging process. 
The second exponential regime gains in importance as the applied voltage increases. 
All these observations can be understood with \cref{eq:PNP_charge_dynamics_late_time}, which predicts that charging goes exponentially with the two timescales of \cref{eq:PNP_charging_time_1,eq:PNP_charging_time_2}.
The second exponential regime goes as $\propto4\sinh^2(\Phi/4)\lambda_D/H\exp(-t/\tau_2)$, whose prefactor explains the absence of the second regime for the smallest potential in \cref{fig:charging_dynamics}(a) and its appearance for larger $\Phi$.
In addition, for $\Phi \leq 1$, the relaxation time $\tau_1$ [\cref{eq:PNP_charging_time_1}] depends only weakly on the applied potential, which results in the same early-time slope of the curves for $\Phi=0.1$ and $1$. 

\Cref{fig:charging_dynamics}(a) shows that charging goes slower for larger applied potentials.
This slowdown can be captured by the \textit{charging time} $t_\text{ch}$, defined as the time at which the pore reaches a certain fraction of its final charge---Ref.~\cite{PRL2014MTL} uses 99\% and, to compare with their results, we make the same choice here.
In \cref{fig:charging_dynamics}(a), $t_\text{ch}$ thus corresponds to the intersection of the charging data with a horizontal line at $\delta Q/Q^f=0.01$---for $\Phi=1$ and $\Phi=4$, we indicate these intersections with stars and the corresponding $t_\text{ch}$ with arrows.
Except for large $\Phi$, the cross-over between the two exponential regimes in \cref{fig:charging_dynamics}(a) is narrow.
$t_\text{ch}$ thus usually falls either in the first or in the second exponential regime.
In \cref{fig:charging_dynamics}(a), $t_\text{ch}$ falls in the first exponential regime for $\Phi=1$ and in the second exponential regime for $\Phi=4$.
We now see that $\Phi$-induced charging slowdown has two different origins.
For small $\Phi$, $t_\text{ch}$ falls in the first exponential regime and increases with $\Phi$ due to the $\cosh(\Phi/2)$ term in the nonlinear $RC$ time $\tau_1$ [\cref{eq:PNP_charging_time_1}].
For large $\Phi$, $t_\text{ch}$ falls in the second exponential regime and increases with $\Phi$ because this regime contains a $\mathcal{O}(\eta)$ prefactor that grows with $\Phi$ [cf.~\cref{eq:PNP_charge_dynamics_late_time}].

To further demonstrate the merits our model, we compare its predictions for $t_\text{ch}(\Phi)$ with corresponding data from direct numerical PNP simulations of Ref.~\cite{PRL2014MTL} of a pore with $H/\lambda_D=20$ \footnote{We deduced $H/\lambda_D=20$, which was not reported in Ref.~\cite{PRL2014MTL}, from Fig.~4 therein.} subject to potentials up to $\Phi=7$.
In \cref{fig:charging_dynamics}(b), the black open circles represent the simulation data of Fig.~5(d) of Ref.~\cite{PRL2014MTL}.
For the mentioned parameters, we cannot use our fully analytical expression \cref{eq:analytic_charging_full} to determine $t_\text{ch}$ as its higher order term $\me^\Phi \lambda_D^2/H^2\approx2.74$ is non-negligible.
We thus use our semi-analytical method---\cref{eq:transport_solve_5_full}, with numerically determined $a_{\pm}$ and $b_{\pm}$.
\Cref{fig:charging_dynamics}(b) shows the charging time $t_\text{ch}(\Phi)$ (blue line) for $H/\lambda_D=20$ 
\footnote{The PNP equations are probably not accurate for the larger $\Phi$ values in \cref{fig:charging_dynamics}(b); we consider $\Phi<\red{8}$ to compare to Ref.~\cite{PRL2014MTL}.}.
All $t_\text{ch}$ data in \cref{fig:charging_dynamics} is scaled by the charging time for small applied potentials $t_0= t_\text{ch}(\Phi\ll1)$.
We estimate $t_0$ with \cref{eq:PNP_charge_dynamics1}: $0.01=(8/\pi^2)\exp{[-\pi^2t_0/(4\tau_{TL})]}$ yields $t_0=\tau_{TL}(4/\pi^2)\ln(800/\pi^2)\approx 1.78 \tau_{TL}$
\footnote{Reference~\cite{PRL2014MTL} scales $t_\text{ch}$ by $\tau_{TL}$.
The $t_\text{ch}/\tau_{TL}$ data in their Fig.~5(d) should approach 1.78 for small applied potential, but it approaches 1, instead.}.
\Cref{fig:charging_dynamics}(b) shows that the prediction from our model---which contains no fitted free parameters---agrees well with the data of Ref.~\cite{PRL2014MTL} for all $\Phi$ considered.
This good agreement is in contrast to Biesheuvel and Bazant's model, whose $t_\text{ch}$ were up to an order of magnitude too large.
Finally, we note that the $t_\text{ch}(\Phi)/t_0$ data in \cref{fig:charging_dynamics}(b) can be approximated as follows.
If 99\% of the charge is reached within the first exponential regime, we can write $0.01=(8/\pi^2)\exp{(-t_\text{ch}/\tau_{1})}$, hence $t_\text{ch}=(4/\pi^2)\tau_{TL}\cosh(\Phi/2)\ln(800/\pi^2)$.
For $\Phi\ll1$, this simplifies to $t_0=(4/\pi^2)\tau_{TL}\ln(800/\pi^2)$. Taking the ratio of these expressions gives $t_\text{ch}/t_0=\cosh(\Phi/2)$.
We see that this approximation describes $t_\text{ch}(\Phi)/t_0$ up to about $\Phi=1$.
Indeed, in \cref{fig:charging_dynamics}(a) we see that, for $\Phi=2$, 99\% of the charge is not reached within the first exponential regime, and the above argument does not hold.


\section{Discussion}\label{sec:discussion}
\subsection{Biexponential decay of two-component systems}
The charging dynamics of our pore is governed by a matrix differential equation \eqref{eq:transport_linear_1}, whose $2\times2$ matrix $\boldsymbol{A}$ has two distinct eigenvalues $\lambda_1$ and $\lambda_2$.
\Cref{eq:appendix_transport_linear_2} shows that these eigenvalues set the relaxation times of the components $g_1$ and $g_2$ of $\boldsymbol{g}=\boldsymbol{P}^{-1} \boldsymbol{\delta\overline{\rho}}$.
The number of timescales (two) in our system is thus a direct consequence of the number of electrolyte components (two).
For $\eta\ll1$, $\boldsymbol{P}=(\boldsymbol{v}_1,\boldsymbol{v}_2)$ takes a simple form, and we find 
\begin{align}\label{eq:smallDukhin}
    \boldsymbol{g}
    &=\frac{1}{2}
    \begin{pmatrix}
        -1 & 1\\
        1 & 1
    \end{pmatrix}
    \begin{pmatrix}
        \overline{\rho}^{0}_+(t,x)-1\\
        \overline{\rho}^{0}_-(t,x)-1
    \end{pmatrix}
    =\frac{1}{2}
    \begin{pmatrix}
        \overline{\rho}^{0}_+-\overline{\rho}^{0}_-\\ \overline{\rho}^{0}_++\overline{\rho}^{0}_--2
    \end{pmatrix}\,.
\end{align}
where we used that $\rho^f_{\pm}\to1$ for $\eta\ll1$.
The elements of $\boldsymbol{g}$ in \cref{eq:smallDukhin} correspond to ionic charge density and salt density.
These quantities thus decouple for $\eta\ll1$ and relax with distinct timescales: $g_1$ relaxes with $\lambda_1$ and $g_2$ relaxes with $\lambda_2$.
For finite $\eta$, the matrix $\boldsymbol{P}$ becomes more complicated and the product $\boldsymbol{P}^{-1}\boldsymbol{\delta\overline{\rho}}$ no-longer yields a charge- and salt-perturbations vector.
Hence, a charge- and salt-perturbations representation no-longer diagonalizes the matrix $\boldsymbol{A}$, which means that salt and charge relaxation become coupled.

The above properties resemble those of electrolyte relaxation between two flat oppositely-charged electrodes \cite{bazant2004diffuse,janssen2018transient}.
When that system is modeled through the PNP equations, the only differences to our setup are in the geometry and its boundary conditions.
For $\Phi\ll1$, the coupled PNP equations for $\rho_+$ and $\rho_-$ again become decoupled in a charge ($\rho_+-\rho_-$) and salt $(\rho_++\rho_-)$ representation.
At $\mathcal{O}(\Phi)$, the salt does not respond, and the charge relaxes with the $RC$ time $\lambda_D L/D$, with $L$ the electrode separation.
The nonlinear charging regime $\Phi\gtrapprox1$ was discussed by Bazant, Thornton, and Ajdori \cite{bazant2004diffuse}.
Through matched asymptotic expansions, they found that the ionic charge density then relaxes biexponentially: the initial $RC$ relaxation is followed at late times by diffusive $L^2/D$ charging.

It would be interesting to study flat-electrode charging through final-state expansions as we did in this paper in \cref{eq:chemical_potential_expansion_2}.
Unfortunately, our approach cannot be transferred directly to the flat-electrodes problem.
In our paper, rather than the PNP equations, we solved the transport equation \eqref{eq:transport_average}, which resulted from the PNP equations after a lubrication approximation ($H\ll L$).
This transport equation only contained the cross-sectional averaged densities $\bar{\rho}_{\pm}^0(t,x)$ and the chemical potential $\mu_{\pm}^0(t,x)$, which did not depend on the $z$-coordinate either.
The absence of $z$-dependence in the transport equation \eqref{eq:transport_average} meant that we could expand the $\mu_{\pm}^0(t,x)$ around the \textit{homogeneous} final state $\bar{\rho}_{\pm}^f$.
This reduced \cref{eq:transport_average} to a matrix differential equation \eqref{eq:transport_linear_1} that was analytically solvable, as its matrix $\boldsymbol{A}$ [\cref{eq:matrix_A_def}] was $t,x,$ and $z$ independent. 
The flat-electrode charging problem is different.
Here, there is no small parameter with which we can reduce the PNP equations to a transport equation in terms of densities averaged in the EDL direction.
Expanding the chemical potentials around the \textit{in-homogeneous} final-state densities then yields a matrix differential equation with a spatially varying matrix.
Such an equation, however, cannot be readily brought by matrix diagonalization to a simple diffusion-type equation like our \cref{eq:appendix_transport_linear_2}.

\subsection{Comparison to Biesheuvel and Bazant \cite{biesheuvel2010nonlinear}}
Biesheuvel and Bazant developed a porous electrode model comprising, at each point in the electrode, a bulk solution in contact with charged double layers \cite{biesheuvel2010nonlinear}.
Specifically, their model accounted for the salt and ionic charge transport through a pore, which exchanged salt and ionic charge with EDLs modeled through Gouy-Chapman theory.
As that is an equilibrium theory, the EDLs of their model were in a quasi-equilibrium that instantaneously adapted to the salt and charge exchange with the quasi-neutral bulk.

Our model has two salient structural similarities to the model in Ref.~\cite{biesheuvel2010nonlinear}:
First, our $\mathcal{O}(1)$ problem \cref{eq:appendix_O-1-eq} describes the equilibrium charge distribution at a cross-section of the pore. 
In Ref.~\cite{biesheuvel2010nonlinear}, this corresponds to their choice to model the EDLs through Gouy-Chapman theory, which is an equilibrium theory.
Second, at $\mathcal{O}(H^2/L^2)$, we found a one-dimensional transport equation [\cref{eq:transport_average}] for the cross-sectional averaged cationic and anionic densities. 
Likewise, Biesheuvel and Bazant use transport equations [Eqs. (9) and (10) there] for the charge and salt adsorption [Eqs. (6) and (8) there].
Again, in their model, the charge and salt adsorption are modeled within the Gouy-Chapman theory, specifically, as integrals of the difference and sum of the densities $\rho_{+}^{f}$ and $\rho_{-}^{f}$ given in \cref{eq:GC_density}. 

Compared to Ref.~\cite{biesheuvel2010nonlinear}, two merits our model are that it is based on a first-principles derivation and that it reproduces the data of Ref.~\cite{PRL2014MTL}.

\subsection{Comparison to Henrique, Zuk, and Gupta \cite{henrique2021charging} and Alizadeh and Mani \cite{alizadeh2017multiscale}}
In Ref.~\cite{henrique2021charging}, Henrique, Zuk, and Gupta studied the charging of a narrow cylindrical pore for arbitrary double layer overlap. 
For thin double layers, their Eq.~(27b) reduces to Posey and Morozumi's expression \cref{eq:TLpotential_difference}.
Like ours, their derivation starts from the PNP equations, but they make two additional assumptions. 

(i) Reference~\cite{henrique2021charging} considered small applied potentials, expanding all observables for small $\Phi$ and accounting only for the $\mathcal{O}(\Phi)$ terms. 
This assumption allowed the authors of Ref.~\cite{henrique2021charging} to render the Poisson equation in radial geometry [Eq.~(15) there] solvable. 
To extend their study to larger $\Phi$, still keeping EDL overlap arbitrary, one should find the cylindrical-pore counterpart of Corkill and Rosenhead's flat-plates Poisson-Boltzmann solution \cite{corkill1939distribution}.
Notice, however, that the validity of their current small-$\Phi$ model is probably governed by a parameter like $\eta$ rather than by $\Phi$: in our work, the second regime of biexponential decay contains the prefactor $\eta=4\exp(\Phi/2)\lambda_D/H$.
As Henrique and coworkers discuss cases of $\lambda_D/H\sim1/2$ (instead of $\le1/20$ as we did here), in practice, the applied potential should be correspondingly smaller to justify ignoring the second exponential regime.

(ii) Next, Ref.~\cite{henrique2021charging} assumed quasi-equilibrium in the radial direction of their cylindrical pore, which they justified by citing Ref.~\cite{alizadeh2017multiscale}.
In Ref.~\cite{alizadeh2017multiscale}, Alizadeh and Mani scaled their electrokinetic equations (14)-(18) by the relevant length scales $L$ and $H$, as we did here.
Taking the limit $H/L\to0$, they found in Eqs. (22)-(28) that their ionic densities were in equilibrium across sections of the pore. 
Finally, they integrated the 3d transport equations over the pore cross section and found reduced 1d equations. 
In spite of the similarity of these steps to our calculations in \cref{sec:appendix_slit-pores}, there is a crucial difference between our methods.
Unlike Ref.~\cite{alizadeh2017multiscale}, we found \textit{asymptotic approximations} to the solutions of the PNP equations \cref{eq:appendix_general_model} (Eqs. (14) and (15) in Ref.~\cite{alizadeh2017multiscale}).
As explained in Ref.~\cite{holmes2012introduction}, performing a scaling analysis to identify a small parameter is one necessary step in this process; plugging in assumed asymptotic expansions for that small parameter [cf.~\cref{eq:appendix_solution_expansion}] into the governing equations is another.
Reference~\cite{alizadeh2017multiscale} did not set the second step. 
Without asymptotic expansions, however, one cannot be sure that the solution of the cross-sectional problem (Eqs.~(22)-(28) in Ref.~\cite{alizadeh2017multiscale}) has the same order in $H/L$ as the variables of the integrated transport (Eqs.~(43) and (44) in Ref.~\cite{alizadeh2017multiscale}). 
We observed that the time-dependent \cref{eq:appendix_O-2_eq} contains both the first and second terms of the asymptotic density and potential expansions over $H/L$.
After cross-sectional averaging of \cref{eq:appendix_O-2_eq}, the second terms of these asymptotic expansions dropped [cf.~\cref{eq:append_boundary}].
Hence, ion transport is governed solely in terms of the first terms of the asymptotic approximations. 
Comparing Eqs. (43) and (44) of Ref.~\cite{alizadeh2017multiscale} (ignoring their fluid velocity term) to our transport equation \eqref{eq:transport_average}, we see that these are actually the same---that is, \textit{if} one reinterpret their densities and chemical potentials as representing the first terms of our asymptotic expansions rather than the full solutions.
Notice that this somewhat trivial result required a nontrivial derivation.


\section{conclusions and outlook}
We have studied the response of an elongated, electrolyte-filled slit pore to a moderate applied potential.
Our approximate analytical solutions to the PNP equations yielded unprecedented insight into the biexponential charging of such pores.
Moreover, we provided the first comprehensive derivation of well-known TL model results.
In our model, we postulated that the ionic density at $x=0$ were instantaneously relaxed [\cref{eq:general_boundary_conditions-bc}].
As shown, this led to results in agreement with prior studies.
Still, in future work, it would be interesting to check by direct numerical simulations to what extend \cref{eq:general_boundary_conditions-bc} agrees with simulated density profiles. 
Related, it would be interesting to extend our model to explicitly account for the electrolyte reservoir with which the pore is in contact.

Future work could also study a case of overlapping EDLs \cite{henrique2021charging}. Instead of the Gouy-Chapman potential \cref{eq:PNP_solution_GC} one should then either use Debye-H\"{u}ckel theory (for $\Phi\ll1$) or the results of Corkill and Rosenhead (for $\Phi\lessapprox2$) \cite{corkill1939distribution}.
Another possible direction is to study nonblocking electrodes, which may shed light on Refs.~\cite{newman1975porous,biesheuvel2011diffuse}.
Last, future work could consider larger applied potentials.
Electrostatic correlations \cite{souza2020continuum} and the finite size of ions \cite{kilic2007stericII} will then become important.
Luckily, substantial parts of \cref{sec:HLexpansion,sec:analytical_model} are actually model-independent and might be directly transferred to study more involved electrolyte model.
One of us---with Sinkov and Akhatov \cite{aslyamov2020relation}---derived precisely the same transport equation \eqref{eq:transport_average} in a DDFT study of confined electrolytes.
Here, we linearized \cref{eq:transport_average} by expanding the chemical potentials [\cref{eq:chemical_potential_expansion_2}] around the final-state ionic densities.
This yielded a linearized transport equation \eqref{eq:transport_linear_1}, whose solution \cref{eq:density_variance_vector} should hold for any system sufficiently close to equilibrium and governed by \cref{eq:transport_average}.
(With increasing potential, a system will move ever further from its initial state; the discarded higher order terms in \cref{eq:chemical_potential_expansion_2} will then become more important.)
The physical properties of a specific pore and electrolyte model enter \cref{eq:density_variance_vector} through the Frobenius covariants, which depend on the expansion coefficients $a_\pm$ and $b_\pm$ [\cref{eq:aibidefinition}].
For PNP, we could determine $a_\pm$ and $b_\pm$ analytically. 
For more involved models, one might need to determine them numerically. 

\section{acknowledgments}
T.A. acknowledges the support from the Russian Science Foundation (project number: 20-72-00183).
We thank Christian Pedersen and Svyatoslav Kondrat for comments on our manuscript and the UiO librarians for providing us with several articles of Daniel-Bekh, Ksenzhek, and Stender.

\appendix
\section{Derivation of \cref{eq:transport_average}}
\label{sec:appendix_slit-pores}
Here, we follow Ref.~\cite{aslyamov2020relation} and derive the transport equation \eqref{eq:transport_average} for long slit pores ($H/L\ll1$). 
First, we change to different dimensionless variables: $\hat{z}=z/H$ and $\hat{x}=x/L$ are the dimensionless $z$- and $x$- coordinates; $\hat{\phi}=\phi/\Phi$ is the scaled dimensionless potential. 
To define dimensionless time and density, we use the $\hat{t}=tD/L^2$ and $\hat{\rho}_{\pm}=H^2\rho_{\pm}/2\Phi \lambda_D^2$, respectively. 
Such variables allow us to explicitly introduce the small parameter $H/L$ into the 2D-PNP \cref{eq:general_model}, as follows
\begin{subequations}
\label{eq:appendix_general_model}
\begin{align}
    \frac{H^2}{L^2}\partial_{\hat{t}} \hat{\rho}_{\pm}&=\frac{H^2}{L^2}\partial_{\hat{x}}(\hat{\rho}_\pm\partial_{\hat{x}}\mu_\pm)+\partial_{\hat{z}}(\hat{\rho}_\pm\partial_{\hat{z}}\mu_\pm),\\
    \frac{H^2}{L^2}\partial_{\hat{x}}^2\hat{\phi}+\partial_{\hat{z}}^2\hat{\phi}&=-(\hat{\rho}_+-\hat{\rho}_-),\\
    \partial_{\hat{z}}\mu_{\pm}(\hat{t},0)&=0,\label{eq:appendix_general_model_zero-flux1}\\
    \partial_{\hat{z}}\mu_{\pm}(\hat{t},1)&=0,\label{eq:appendix_general_model_zero-flux2}\\
    \hat{\phi}(\hat{t},0)&=1,\\
    \partial_{\hat{z}}\hat{\phi}(\hat{t},1/2)&=0,
\end{align}
\end{subequations}
where \cref{eq:appendix_general_model_zero-flux1,eq:appendix_general_model_zero-flux2} express the condition of the zero ionic flux through the pore walls.

As \cref{eq:appendix_general_model} contains only even powers (zero and two) of the small parameter $H/L$, we seek solutions $\hat{\rho}_\pm$ and $\hat{\phi}$ to \cref{eq:appendix_general_model} in terms of series with even powers of $H/L$ too,
\begin{subequations}
\label{eq:appendix_solution_expansion}
\begin{align}
\hat{\rho}_\pm &= \hat{\rho}^0_\pm + \frac{H^2}{L^2} \hat{\rho}^1_\pm + \mathcal{O}\left(\frac{H^{4}}{L^{4}}\right), \\
\hat{\phi} &= \hat{\phi}^0 + \frac{H^2}{L^2} \hat{\phi}^1 + \mathcal{O}\left(\frac{H^4}{L^4}\right).
\end{align}
\end{subequations}
We insert \cref{eq:appendix_solution_expansion} into \cref{eq:appendix_general_model} and collect terms of the same order in $H/L$.
At $\mathcal{O}(1)$, we find
\begin{subequations}
\label{eq:appendix_O-1}
\begin{align}
\label{eq:appendix_O-1-eq}
    0&=\partial_{\hat{z}}\hat{\rho}_\pm^0\partial_{\hat{z}}\mu_\pm^0,\\
    \label{eq:appendix_O-1-Poisson}
    \partial_{\hat{z}}^2\hat{\phi}^0&=-(\hat{\rho}_+^0-\hat{\rho}_-^0),\\
    \partial_{\hat{z}}\mu^0_{\pm}(\hat{t},0)&=0,\label{eq:appendix_O-1_cp-bc1}\\
    \partial_{\hat{z}}\mu^0_{\pm}(\hat{t},1)&=0,\label{eq:appendix_O-1_cp-bc2}\\
    \hat{\phi}^0(\hat{t},0)&=1,\label{eq:appendix_O-1-Poisson-bc1}\\
    \partial_{\hat{z}}\hat{\phi}^0(\hat{t},1/2)&=0,\label{eq:appendix_O-1-Poisson-bc2}
\end{align}
\end{subequations}
where $\mu_\pm^0=\mu_\pm(\hat{\rho}_\pm^0,\hat{\phi}^0)$ is the $\mathcal{O}(1)$ chemical potential. 
Notice that the $\mathcal{O}(1)$-problem does not depend on time. 
From \cref{eq:appendix_O-1-eq,eq:appendix_O-1_cp-bc1,eq:appendix_O-1_cp-bc2}, we see that the chemical potential does not depend on $\hat{z}$-coordinate $\mu_\pm^0=\mu_\pm^0(\hat{t},\hat{x})$. 
This condition means that the $\mathcal{O}(1)$ density distributions can be found from \cref{eq:general_model_mu}, which results in \cref{eq:density_distribution}. 
The remaining \cref{eq:appendix_O-1-Poisson,eq:appendix_O-1-Poisson-bc1,eq:appendix_O-1-Poisson-bc2} give us the $\mathcal{O}(1)$-Poisson equation \eqref{eq:Poisson_equation}, which can be solved numerically and analytically (see \cref{sec:analytical_model,sec:appendix_numerical}, respectively).

Inserting \cref{eq:appendix_solution_expansion} into \cref{eq:appendix_general_model} gives, at $\mathcal{O}(H^2/L^2)$,
\begin{subequations}
\label{eq:appendix_O-2}
\begin{align}
    \label{eq:appendix_O-2_eq}
    \partial_{\hat{t}} \hat{\rho}_\pm^0&=\partial_{\hat{x}}\left(\hat{\rho}_\pm^0\partial_{\hat{x}}\mu_\pm^0\right)+\partial_{\hat{z}}\left(\hat{\rho}_\pm^1\partial_{\hat{z}}\mu_\pm^0+\hat{\rho}_\pm^0\partial_{\hat{z}}\mu_\pm^1\right),\\
    \label{eq:appendix_O-2_mu-1}
    \mu_\pm^1&=\frac{\delta\mu_\pm}{\delta\hat{\rho}_+}\hat{\rho}_+^1+\frac{\delta\mu_\pm}{\delta\hat{\rho}_-}\hat{\rho}_-^1+\frac{\delta\mu_\pm}{\delta \hat{\phi}}\hat{\phi}^1,\\
    \partial_{\hat{z}}\mu^1_{\pm}(\hat{t},0)&=0,\label{eq:appendix_O-2_cp-bc1}\\
    \partial_{\hat{z}}\mu^1_{\pm}(\hat{t},1)&=0,\label{eq:appendix_O-2_cp-bc2}\\
    \hat{\phi}^1(\hat{t},0)&=0,\\
    \partial_{\hat{z}}\hat{\phi}^1(\hat{t},1/2)&=0,
\end{align}
\end{subequations}
where the derivatives in \cref{eq:appendix_O-2_mu-1} are calculated at $\hat{\rho}_\pm=\hat{\rho}_\pm^0$ and $\hat{\phi}=\hat{\phi}^0$. 

We integrate \cref{eq:appendix_O-2_eq} over $z$ from $0$ to $1$ and find
\begin{equation}
    \label{eq:appendix_transport_equation}
    \partial_{\hat{t}} \overline{\hat{\rho}}_\pm^0=\partial_{\hat{x}}\left(\int_0^1d\hat{z}\hat{\rho}_\pm^0\partial_{\hat{x}}\mu_\pm^0(\hat{t},\hat{x})\right)=\partial_{\hat{x}}\left(\overline{\hat{\rho}}_\pm^0\partial_{\hat{x}}\mu_\pm^0\right),
\end{equation}
where, for the first equality, we used 
\begin{equation}\label{eq:append_boundary}
\left.(\hat{\rho}_\pm^1\partial_{\hat{z}}\mu_\pm^0+\hat{\rho}_\pm^0\partial_{\hat{z}}\mu_\pm^1)\right|_{\hat{z}=0}^1=0\,,
\end{equation}
which follows from the conditions $\partial_{\hat{x}}\mu_\pm^0=\partial_{\hat{x}}\mu_\pm^1=0$ for $\hat{z}=0$ and $\hat{z}=1$, see \cref{eq:appendix_O-1_cp-bc1,eq:appendix_O-1_cp-bc2,eq:appendix_O-2_cp-bc1,eq:appendix_O-2_cp-bc2}.
For the second equality in \cref{eq:appendix_transport_equation}, we used that $\mu_\pm^0(x,t)$ does not depend on $z$, which follows from \cref{eq:appendix_O-1-eq,eq:appendix_O-1_cp-bc1,eq:appendix_O-1_cp-bc2}, and can thus be taken out of the integral. 
Returning to the variables of the main text, we arrive at \cref{eq:transport_average}.


\section{Numerical calculations}
\label{sec:appendix_numerical}
We numerically solve \cref{eq:transport_average} through two sub-tasks:
(i) the calculation of the chemical potentials data to obtain the functions of two variables $\mu_\pm(\overline{\rho}_+,\overline{\rho}_-)$ by interpolation;
(ii) the solution of the transport equation \eqref{eq:transport_average} for given functions $\mu_\pm(\overline{\rho}_+,\overline{\rho}_-)$.

Sub-task (i) corresponds to finding a self-consistent solution of \cref{eq:normalized_Poisson,eq:norm}. These equations calculated for densities from a discrete two-dimensional set $\{\overline{\rho}^{(n)}_+\}_{n=0}^{N}\times\{\overline{\rho}^{(n)}_-\}_{n=0}^{N}$, which is the Cartesian product of the one-dimensional lists containing the following elements for $\Phi>0$:
\begin{subequations}
\label{eq:appendix_numerical_density_grid}
\begin{align}
    \rho^{(n)}_+&=\overline{\rho}^f_+-\epsilon_++\frac{1-\overline{\rho}^f_++2\epsilon_+}{N}n, \quad \text{for} \quad n=0,\dots, N, \\
    \rho^{(n)}_-&=1-\epsilon_-+\frac{\overline{\rho}^f_--1+2\epsilon_-}{N}n, \quad \text{for} \quad n=0,\dots, N,
\end{align}
\end{subequations}
with $N$ the number of elements/gridpoint in our density discretization, $\epsilon_\pm=2|1-\rho_\pm|/N$ is a parameter which extends the data-set beyond the range $(\overline{\rho}^f_+,\overline{\rho}^f_-)$. In our calculations we used $N=10$.
We use the Python library SciPy to solve \cref{eq:normalized_Poisson,eq:norm} for the densities \cref{eq:appendix_numerical_density_grid}. 
Then, we interpolate the calculated data using the standard interpolation function of Wolfram Mathematica, which gives us the functions $\mu_\pm^\text{int}(\overline{\rho}_+,\overline{\rho}_-)$. 
We use these functions to numerically determine the coefficients $a_\pm$ and $b_\pm$, as follows
\begin{equation}\label{eq:aibidefinition-numerical}
    a_\pm=\left.\frac{ \partial\mu^\text{int}_\pm}{\partial\overline{\rho}_+^0}\right|_{\overline{\rho}^f_+}, \qquad \left. 
    b_\pm=\frac{ \partial\mu^\text{int}_\pm}{\partial\overline{\rho}_-^0}\right|_{\overline{\rho}^f_-}.
\end{equation}

To solve sub-task (ii), we follow Ref.~\cite{aslyamov2020relation}: spatial discretization along $x$-coordinate is performed on a uniform staggered grid using finite volume method; 
the resulting system of the ODEs is solved with the built-in method of Wolfram Mathematica. 


%


\begin{thebibliography}{47}%
\makeatletter
\providecommand \@ifxundefined [1]{%
 \@ifx{#1\undefined}
}%
\providecommand \@ifnum [1]{%
 \ifnum #1\expandafter \@firstoftwo
 \else \expandafter \@secondoftwo
 \fi
}%
\providecommand \@ifx [1]{%
 \ifx #1\expandafter \@firstoftwo
 \else \expandafter \@secondoftwo
 \fi
}%
\providecommand \natexlab [1]{#1}%
\providecommand \enquote  [1]{``#1''}%
\providecommand \bibnamefont  [1]{#1}%
\providecommand \bibfnamefont [1]{#1}%
\providecommand \citenamefont [1]{#1}%
\providecommand \href@noop [0]{\@secondoftwo}%
\providecommand \href [0]{\begingroup \@sanitize@url \@href}%
\providecommand \@href[1]{\@@startlink{#1}\@@href}%
\providecommand \@@href[1]{\endgroup#1\@@endlink}%
\providecommand \@sanitize@url [0]{\catcode `\\12\catcode `\$12\catcode
  `\&12\catcode `\#12\catcode `\^12\catcode `\_12\catcode `\%12\relax}%
\providecommand \@@startlink[1]{}%
\providecommand \@@endlink[0]{}%
\providecommand \url  [0]{\begingroup\@sanitize@url \@url }%
\providecommand \@url [1]{\endgroup\@href {#1}{\urlprefix }}%
\providecommand \urlprefix  [0]{URL }%
\providecommand \Eprint [0]{\href }%
\providecommand \doibase [0]{https://doi.org/}%
\providecommand \selectlanguage [0]{\@gobble}%
\providecommand \bibinfo  [0]{\@secondoftwo}%
\providecommand \bibfield  [0]{\@secondoftwo}%
\providecommand \translation [1]{[#1]}%
\providecommand \BibitemOpen [0]{}%
\providecommand \bibitemStop [0]{}%
\providecommand \bibitemNoStop [0]{.\EOS\space}%
\providecommand \EOS [0]{\spacefactor3000\relax}%
\providecommand \BibitemShut  [1]{\csname bibitem#1\endcsname}%
\let\auto@bib@innerbib\@empty
\bibitem [{\citenamefont {Lasia}(2014)}]{lasia2002electrochemical}%
  \BibitemOpen
  \bibfield  {author} {\bibinfo {author} {\bibfnamefont {A.}~\bibnamefont
  {Lasia}},\ }\href {https://doi.org/10.1007/978-1-4614-8933-7} {\emph
  {\bibinfo {title} {Electrochemical impedance spectroscopy and its
  applications}}}\ (\bibinfo  {publisher} {Springer},\ \bibinfo {year}
  {2014})\BibitemShut {NoStop}%
\bibitem [{\citenamefont {Huang}\ \emph {et~al.}(2020)\citenamefont {Huang},
  \citenamefont {Gao}, \citenamefont {Luo}, \citenamefont {Wang}, \citenamefont
  {Li}, \citenamefont {Chen},\ and\ \citenamefont {Zhang}}]{huang2020review}%
  \BibitemOpen
  \bibfield  {author} {\bibinfo {author} {\bibfnamefont {J.}~\bibnamefont
  {Huang}}, \bibinfo {author} {\bibfnamefont {Y.}~\bibnamefont {Gao}}, \bibinfo
  {author} {\bibfnamefont {J.}~\bibnamefont {Luo}}, \bibinfo {author}
  {\bibfnamefont {S.}~\bibnamefont {Wang}}, \bibinfo {author} {\bibfnamefont
  {C.}~\bibnamefont {Li}}, \bibinfo {author} {\bibfnamefont {S.}~\bibnamefont
  {Chen}},\ and\ \bibinfo {author} {\bibfnamefont {J.}~\bibnamefont {Zhang}},\
  }\href {https://doi.org/10.1149/1945-7111/abc655} {\bibfield  {journal}
  {\bibinfo  {journal} {J. Electrochem. Soc.}\ }\textbf {\bibinfo {volume}
  {167}},\ \bibinfo {pages} {166503} (\bibinfo {year} {2020})}\BibitemShut
  {NoStop}%
\bibitem [{\citenamefont {Conway}(2013)}]{conway2013electrochemical}%
  \BibitemOpen
  \bibfield  {author} {\bibinfo {author} {\bibfnamefont {B.~E.}\ \bibnamefont
  {Conway}},\ }\href@noop {} {\emph {\bibinfo {title} {Electrochemical
  supercapacitors: scientific fundamentals and technological applications}}}\
  (\bibinfo  {publisher} {Springer Science \& Business Media},\ \bibinfo {year}
  {2013})\BibitemShut {NoStop}%
\bibitem [{\citenamefont {P{\'e}an}\ \emph {et~al.}(2014)\citenamefont
  {P{\'e}an}, \citenamefont {Merlet}, \citenamefont {Rotenberg}, \citenamefont
  {Madden}, \citenamefont {Taberna}, \citenamefont {Daffos}, \citenamefont
  {Salanne},\ and\ \citenamefont {Simon}}]{pean2014dynamics}%
  \BibitemOpen
  \bibfield  {author} {\bibinfo {author} {\bibfnamefont {C.}~\bibnamefont
  {P{\'e}an}}, \bibinfo {author} {\bibfnamefont {C.}~\bibnamefont {Merlet}},
  \bibinfo {author} {\bibfnamefont {B.}~\bibnamefont {Rotenberg}}, \bibinfo
  {author} {\bibfnamefont {P.~A.}\ \bibnamefont {Madden}}, \bibinfo {author}
  {\bibfnamefont {P.-L.}\ \bibnamefont {Taberna}}, \bibinfo {author}
  {\bibfnamefont {B.}~\bibnamefont {Daffos}}, \bibinfo {author} {\bibfnamefont
  {M.}~\bibnamefont {Salanne}},\ and\ \bibinfo {author} {\bibfnamefont
  {P.}~\bibnamefont {Simon}},\ }\href {https://doi.org/10.1021/nn4058243}
  {\bibfield  {journal} {\bibinfo  {journal} {ACS Nano}\ }\textbf {\bibinfo
  {volume} {8}},\ \bibinfo {pages} {1576} (\bibinfo {year} {2014})}\BibitemShut
  {NoStop}%
\bibitem [{\citenamefont {Kondrat}\ \emph {et~al.}(2014)\citenamefont
  {Kondrat}, \citenamefont {Wu}, \citenamefont {Qiao},\ and\ \citenamefont
  {Kornyshev}}]{kondrat2014NM}%
  \BibitemOpen
  \bibfield  {author} {\bibinfo {author} {\bibfnamefont {S.}~\bibnamefont
  {Kondrat}}, \bibinfo {author} {\bibfnamefont {P.}~\bibnamefont {Wu}},
  \bibinfo {author} {\bibfnamefont {R.}~\bibnamefont {Qiao}},\ and\ \bibinfo
  {author} {\bibfnamefont {A.~A.}\ \bibnamefont {Kornyshev}},\ }\href
  {https://doi.org/10.1038/nmat3916} {\bibfield  {journal} {\bibinfo  {journal}
  {Nat. Mater.}\ }\textbf {\bibinfo {volume} {13}},\ \bibinfo {pages} {387}
  (\bibinfo {year} {2014})}\BibitemShut {NoStop}%
\bibitem [{\citenamefont {Pak}\ and\ \citenamefont
  {Hwang}(2016)}]{pak2016charging}%
  \BibitemOpen
  \bibfield  {author} {\bibinfo {author} {\bibfnamefont {A.~J.}\ \bibnamefont
  {Pak}}\ and\ \bibinfo {author} {\bibfnamefont {G.~S.}\ \bibnamefont
  {Hwang}},\ }\href {https://doi.org/10.1021/acs.jpcc.6b06637} {\bibfield
  {journal} {\bibinfo  {journal} {J. Phys. Chem. C}\ }\textbf {\bibinfo
  {volume} {120}},\ \bibinfo {pages} {24560} (\bibinfo {year}
  {2016})}\BibitemShut {NoStop}%
\bibitem [{\citenamefont {He}\ \emph {et~al.}(2016)\citenamefont {He},
  \citenamefont {Qiao}, \citenamefont {Vatamanu}, \citenamefont {Borodin},
  \citenamefont {Bedrov}, \citenamefont {Huang},\ and\ \citenamefont
  {Sumpter}}]{he2016JPCL}%
  \BibitemOpen
  \bibfield  {author} {\bibinfo {author} {\bibfnamefont {Y.}~\bibnamefont
  {He}}, \bibinfo {author} {\bibfnamefont {R.}~\bibnamefont {Qiao}}, \bibinfo
  {author} {\bibfnamefont {J.}~\bibnamefont {Vatamanu}}, \bibinfo {author}
  {\bibfnamefont {O.}~\bibnamefont {Borodin}}, \bibinfo {author} {\bibfnamefont
  {D.}~\bibnamefont {Bedrov}}, \bibinfo {author} {\bibfnamefont
  {J.}~\bibnamefont {Huang}},\ and\ \bibinfo {author} {\bibfnamefont {B.~G.}\
  \bibnamefont {Sumpter}},\ }\href
  {https://doi.org/10.1021/acs.jpclett.5b02378} {\bibfield  {journal} {\bibinfo
   {journal} {J. Phys. Chem. Lett.}\ }\textbf {\bibinfo {volume} {7}},\
  \bibinfo {pages} {36} (\bibinfo {year} {2016})}\BibitemShut {NoStop}%
\bibitem [{\citenamefont {Breitsprecher}\ \emph {et~al.}(2017)\citenamefont
  {Breitsprecher}, \citenamefont {Abele}, \citenamefont {Kondrat},\ and\
  \citenamefont {Holm}}]{breitsprecher_effect_2017}%
  \BibitemOpen
  \bibfield  {author} {\bibinfo {author} {\bibfnamefont {K.}~\bibnamefont
  {Breitsprecher}}, \bibinfo {author} {\bibfnamefont {M.}~\bibnamefont
  {Abele}}, \bibinfo {author} {\bibfnamefont {S.}~\bibnamefont {Kondrat}},\
  and\ \bibinfo {author} {\bibfnamefont {C.}~\bibnamefont {Holm}},\ }\href
  {https://doi.org/10.1063/1.4986346} {\bibfield  {journal} {\bibinfo
  {journal} {J. Chem. Phys.}\ }\textbf {\bibinfo {volume} {147}},\ \bibinfo
  {pages} {104708} (\bibinfo {year} {2017})}\BibitemShut {NoStop}%
\bibitem [{\citenamefont {Breitsprecher}\ \emph {et~al.}(2018)\citenamefont
  {Breitsprecher}, \citenamefont {Holm},\ and\ \citenamefont
  {Kondrat}}]{breitsprecher2018charge}%
  \BibitemOpen
  \bibfield  {author} {\bibinfo {author} {\bibfnamefont {K.}~\bibnamefont
  {Breitsprecher}}, \bibinfo {author} {\bibfnamefont {C.}~\bibnamefont
  {Holm}},\ and\ \bibinfo {author} {\bibfnamefont {S.}~\bibnamefont
  {Kondrat}},\ }\href {https://doi.org/10.1021/acsnano.8b04785} {\bibfield
  {journal} {\bibinfo  {journal} {ACS Nano}\ }\textbf {\bibinfo {volume}
  {12}},\ \bibinfo {pages} {9733} (\bibinfo {year} {2018})}\BibitemShut
  {NoStop}%
\bibitem [{\citenamefont {Breitsprecher}\ \emph {et~al.}(2020)\citenamefont
  {Breitsprecher}, \citenamefont {Janssen}, \citenamefont {Srimuk},
  \citenamefont {Mehdi}, \citenamefont {Presser}, \citenamefont {Holm},\ and\
  \citenamefont {Kondrat}}]{breitsprecher2020speed}%
  \BibitemOpen
  \bibfield  {author} {\bibinfo {author} {\bibfnamefont {K.}~\bibnamefont
  {Breitsprecher}}, \bibinfo {author} {\bibfnamefont {M.}~\bibnamefont
  {Janssen}}, \bibinfo {author} {\bibfnamefont {P.}~\bibnamefont {Srimuk}},
  \bibinfo {author} {\bibfnamefont {B.~L.}\ \bibnamefont {Mehdi}}, \bibinfo
  {author} {\bibfnamefont {V.}~\bibnamefont {Presser}}, \bibinfo {author}
  {\bibfnamefont {C.}~\bibnamefont {Holm}},\ and\ \bibinfo {author}
  {\bibfnamefont {S.}~\bibnamefont {Kondrat}},\ }\href
  {https://doi.org/10.1038/s41467-020-19903-6} {\bibfield  {journal} {\bibinfo
  {journal} {Nat. Commun.}\ }\textbf {\bibinfo {volume} {11}},\ \bibinfo
  {pages} {1} (\bibinfo {year} {2020})}\BibitemShut {NoStop}%
\bibitem [{\citenamefont {Bi}\ \emph {et~al.}(2020)\citenamefont {Bi},
  \citenamefont {Banda}, \citenamefont {Chen}, \citenamefont {Niu},
  \citenamefont {Chen}, \citenamefont {Wu}, \citenamefont {Wang}, \citenamefont
  {Wang}, \citenamefont {Feng}, \citenamefont {Chen} \emph
  {et~al.}}]{bi2020molecular}%
  \BibitemOpen
  \bibfield  {author} {\bibinfo {author} {\bibfnamefont {S.}~\bibnamefont
  {Bi}}, \bibinfo {author} {\bibfnamefont {H.}~\bibnamefont {Banda}}, \bibinfo
  {author} {\bibfnamefont {M.}~\bibnamefont {Chen}}, \bibinfo {author}
  {\bibfnamefont {L.}~\bibnamefont {Niu}}, \bibinfo {author} {\bibfnamefont
  {M.}~\bibnamefont {Chen}}, \bibinfo {author} {\bibfnamefont {T.}~\bibnamefont
  {Wu}}, \bibinfo {author} {\bibfnamefont {J.}~\bibnamefont {Wang}}, \bibinfo
  {author} {\bibfnamefont {R.}~\bibnamefont {Wang}}, \bibinfo {author}
  {\bibfnamefont {J.}~\bibnamefont {Feng}}, \bibinfo {author} {\bibfnamefont
  {T.}~\bibnamefont {Chen}}, \emph {et~al.},\ }\href
  {https://doi.org/10.1038/s41563-019-0598-7} {\bibfield  {journal} {\bibinfo
  {journal} {Nat. Mater.}\ }\textbf {\bibinfo {volume} {19}},\ \bibinfo {pages}
  {552} (\bibinfo {year} {2020})}\BibitemShut {NoStop}%
\bibitem [{\citenamefont {Mo}\ \emph {et~al.}(2020)\citenamefont {Mo},
  \citenamefont {Bi}, \citenamefont {Zhang}, \citenamefont {Presser},
  \citenamefont {Wang}, \citenamefont {Gogotsi},\ and\ \citenamefont
  {Feng}}]{mo2020ion}%
  \BibitemOpen
  \bibfield  {author} {\bibinfo {author} {\bibfnamefont {T.}~\bibnamefont
  {Mo}}, \bibinfo {author} {\bibfnamefont {S.}~\bibnamefont {Bi}}, \bibinfo
  {author} {\bibfnamefont {Y.}~\bibnamefont {Zhang}}, \bibinfo {author}
  {\bibfnamefont {V.}~\bibnamefont {Presser}}, \bibinfo {author} {\bibfnamefont
  {X.}~\bibnamefont {Wang}}, \bibinfo {author} {\bibfnamefont {Y.}~\bibnamefont
  {Gogotsi}},\ and\ \bibinfo {author} {\bibfnamefont {G.}~\bibnamefont
  {Feng}},\ }\href {https://doi.org/10.1021/acsnano.9b09648} {\bibfield
  {journal} {\bibinfo  {journal} {ACS Nano}\ }\textbf {\bibinfo {volume}
  {14}},\ \bibinfo {pages} {2395} (\bibinfo {year} {2020})}\BibitemShut
  {NoStop}%
\bibitem [{\citenamefont {Lian}\ \emph {et~al.}(2020)\citenamefont {Lian},
  \citenamefont {Janssen}, \citenamefont {Liu},\ and\ \citenamefont {van
  Roij}}]{lian2020blessing}%
  \BibitemOpen
  \bibfield  {author} {\bibinfo {author} {\bibfnamefont {C.}~\bibnamefont
  {Lian}}, \bibinfo {author} {\bibfnamefont {M.}~\bibnamefont {Janssen}},
  \bibinfo {author} {\bibfnamefont {H.}~\bibnamefont {Liu}},\ and\ \bibinfo
  {author} {\bibfnamefont {R.}~\bibnamefont {van Roij}},\ }\href
  {https://doi.org/10.1103/PhysRevLett.124.076001} {\bibfield  {journal}
  {\bibinfo  {journal} {Phys. Rev. Lett.}\ }\textbf {\bibinfo {volume} {124}},\
  \bibinfo {pages} {076001} (\bibinfo {year} {2020})}\BibitemShut {NoStop}%
\bibitem [{\citenamefont {Sakaguchi}\ and\ \citenamefont
  {Baba}(2007)}]{sakaguchi2007}%
  \BibitemOpen
  \bibfield  {author} {\bibinfo {author} {\bibfnamefont {H.}~\bibnamefont
  {Sakaguchi}}\ and\ \bibinfo {author} {\bibfnamefont {R.}~\bibnamefont
  {Baba}},\ }\href {https://doi.org/10.1103/PhysRevE.76.011501} {\bibfield
  {journal} {\bibinfo  {journal} {Phys. Rev. E}\ }\textbf {\bibinfo {volume}
  {76}},\ \bibinfo {pages} {011501} (\bibinfo {year} {2007})}\BibitemShut
  {NoStop}%
\bibitem [{\citenamefont {Lim}\ \emph {et~al.}(2009)\citenamefont {Lim},
  \citenamefont {Whitcomb}, \citenamefont {Boyd},\ and\ \citenamefont
  {Varghese}}]{lim2009effect}%
  \BibitemOpen
  \bibfield  {author} {\bibinfo {author} {\bibfnamefont {J.}~\bibnamefont
  {Lim}}, \bibinfo {author} {\bibfnamefont {J.~D.}\ \bibnamefont {Whitcomb}},
  \bibinfo {author} {\bibfnamefont {J.~G.}\ \bibnamefont {Boyd}},\ and\
  \bibinfo {author} {\bibfnamefont {J.}~\bibnamefont {Varghese}},\ }\href
  {https://doi.org/10.1007/s00466-008-0322-y} {\bibfield  {journal} {\bibinfo
  {journal} {Comput. Mech.}\ }\textbf {\bibinfo {volume} {43}},\ \bibinfo
  {pages} {461} (\bibinfo {year} {2009})}\BibitemShut {NoStop}%
\bibitem [{\citenamefont {Mirzadeh}\ and\ \citenamefont
  {Gibou}(2014)}]{MIRZADEH2014633}%
  \BibitemOpen
  \bibfield  {author} {\bibinfo {author} {\bibfnamefont {M.}~\bibnamefont
  {Mirzadeh}}\ and\ \bibinfo {author} {\bibfnamefont {F.}~\bibnamefont
  {Gibou}},\ }\href {https://doi.org/https://doi.org/10.1016/j.jcp.2014.06.039}
  {\bibfield  {journal} {\bibinfo  {journal} {J. Comput. Phys.}\ }\textbf
  {\bibinfo {volume} {274}},\ \bibinfo {pages} {633} (\bibinfo {year}
  {2014})}\BibitemShut {NoStop}%
\bibitem [{\citenamefont {Mirzadeh}\ \emph {et~al.}(2014)\citenamefont
  {Mirzadeh}, \citenamefont {Gibou},\ and\ \citenamefont
  {Squires}}]{PRL2014MTL}%
  \BibitemOpen
  \bibfield  {author} {\bibinfo {author} {\bibfnamefont {M.}~\bibnamefont
  {Mirzadeh}}, \bibinfo {author} {\bibfnamefont {F.}~\bibnamefont {Gibou}},\
  and\ \bibinfo {author} {\bibfnamefont {T.~M.}\ \bibnamefont {Squires}},\
  }\href {https://doi.org/10.1103/PhysRevLett.113.097701} {\bibfield  {journal}
  {\bibinfo  {journal} {Phys. Rev. Lett.}\ }\textbf {\bibinfo {volume} {113}},\
  \bibinfo {pages} {097701} (\bibinfo {year} {2014})}\BibitemShut {NoStop}%
\bibitem [{\citenamefont {Henrique}\ \emph {et~al.}(2021)\citenamefont
  {Henrique}, \citenamefont {Zuk},\ and\ \citenamefont
  {Gupta}}]{henrique2021charging}%
  \BibitemOpen
  \bibfield  {author} {\bibinfo {author} {\bibfnamefont {F.}~\bibnamefont
  {Henrique}}, \bibinfo {author} {\bibfnamefont {P.~J.}\ \bibnamefont {Zuk}},\
  and\ \bibinfo {author} {\bibfnamefont {A.}~\bibnamefont {Gupta}},\ }\href
  {https://doi.org/10.1039/D1SM01239H} {\bibfield  {journal} {\bibinfo
  {journal} {Soft Matter}\ }\textbf {\bibinfo {volume} {18}},\ \bibinfo {pages}
  {198} (\bibinfo {year} {2021})}\BibitemShut {NoStop}%
\bibitem [{\citenamefont {Aslyamov}\ \emph {et~al.}(2020)\citenamefont
  {Aslyamov}, \citenamefont {Sinkov},\ and\ \citenamefont
  {Akhatov}}]{aslyamov2020relation}%
  \BibitemOpen
  \bibfield  {author} {\bibinfo {author} {\bibfnamefont {T.}~\bibnamefont
  {Aslyamov}}, \bibinfo {author} {\bibfnamefont {K.}~\bibnamefont {Sinkov}},\
  and\ \bibinfo {author} {\bibfnamefont {I.}~\bibnamefont {Akhatov}},\
  }\href@noop {} {\bibinfo {title} {Relation between charging times and storage
  properties of nanoporous supercapacitors}} (\bibinfo {year} {2020}),\ \Eprint
  {https://arxiv.org/abs/2011.04575} {arXiv:2011.04575 [cond-mat.stat-mech]}
  \BibitemShut {NoStop}%
\bibitem [{\citenamefont {Tomlin}\ \emph {et~al.}(2021)\citenamefont {Tomlin},
  \citenamefont {Roy}, \citenamefont {Kirk}, \citenamefont {Marinescu},\ and\
  \citenamefont {Gillespie}}]{tomlin2021impedance}%
  \BibitemOpen
  \bibfield  {author} {\bibinfo {author} {\bibfnamefont {R.~J.}\ \bibnamefont
  {Tomlin}}, \bibinfo {author} {\bibfnamefont {T.}~\bibnamefont {Roy}},
  \bibinfo {author} {\bibfnamefont {T.~L.}\ \bibnamefont {Kirk}}, \bibinfo
  {author} {\bibfnamefont {M.}~\bibnamefont {Marinescu}},\ and\ \bibinfo
  {author} {\bibfnamefont {D.}~\bibnamefont {Gillespie}},\ }\href@noop {}
  {\bibinfo {title} {Impedance response of ionic liquids in long slit pores}}
  (\bibinfo {year} {2021}),\ \Eprint {https://arxiv.org/abs/2110.07014}
  {arXiv:2110.07014 [cond-mat.soft]} \BibitemShut {NoStop}%
\bibitem [{\citenamefont {Biesheuvel}\ and\ \citenamefont
  {Bazant}(2010)}]{biesheuvel2010nonlinear}%
  \BibitemOpen
  \bibfield  {author} {\bibinfo {author} {\bibfnamefont {P.~M.}\ \bibnamefont
  {Biesheuvel}}\ and\ \bibinfo {author} {\bibfnamefont {M.~Z.}\ \bibnamefont
  {Bazant}},\ }\href {https://doi.org/10.1103/PhysRevE.81.031502} {\bibfield
  {journal} {\bibinfo  {journal} {Phys. Rev. E}\ }\textbf {\bibinfo {volume}
  {81}},\ \bibinfo {pages} {031502} (\bibinfo {year} {2010})}\BibitemShut
  {NoStop}%
\bibitem [{\citenamefont {Daniel-Bekh}(1948)}]{danielbek1948}%
  \BibitemOpen
  \bibfield  {author} {\bibinfo {author} {\bibfnamefont {V.~S.}\ \bibnamefont
  {Daniel-Bekh}},\ }\href@noop {} {\bibfield  {journal} {\bibinfo  {journal}
  {Zh. Fiz. Khim. SSR}\ }\textbf {\bibinfo {volume} {22}},\ \bibinfo {pages}
  {697} (\bibinfo {year} {1948})}\BibitemShut {NoStop}%
\bibitem [{\citenamefont {Ksenzhek}\ and\ \citenamefont
  {Stender}(1956)}]{ksenzhek1956}%
  \BibitemOpen
  \bibfield  {author} {\bibinfo {author} {\bibfnamefont {O.~S.}\ \bibnamefont
  {Ksenzhek}}\ and\ \bibinfo {author} {\bibfnamefont {V.~V.}\ \bibnamefont
  {Stender}},\ }\href@noop {} {\bibfield  {journal} {\bibinfo  {journal} {Dokl.
  Akad. Nauk SSSR}\ }\textbf {\bibinfo {volume} {106}},\ \bibinfo {pages} {487}
  (\bibinfo {year} {1956})}\BibitemShut {NoStop}%
\bibitem [{\citenamefont {de~Levie}(1963)}]{levie1963}%
  \BibitemOpen
  \bibfield  {author} {\bibinfo {author} {\bibfnamefont {R.}~\bibnamefont
  {de~Levie}},\ }\href
  {https://doi.org/https://doi.org/10.1016/0013-4686(63)80042-0} {\bibfield
  {journal} {\bibinfo  {journal} {Electrochim. Acta}\ }\textbf {\bibinfo
  {volume} {8}},\ \bibinfo {pages} {751 } (\bibinfo {year} {1963})}\BibitemShut
  {NoStop}%
\bibitem [{\citenamefont {de~Levie}(1964)}]{levie1964porous}%
  \BibitemOpen
  \bibfield  {author} {\bibinfo {author} {\bibfnamefont {R.}~\bibnamefont
  {de~Levie}},\ }\href {https://doi.org/10.1016/0013-4686(64)85015-5}
  {\bibfield  {journal} {\bibinfo  {journal} {Electrochim. Acta}\ }\textbf
  {\bibinfo {volume} {9}},\ \bibinfo {pages} {1231} (\bibinfo {year}
  {1964})}\BibitemShut {NoStop}%
\bibitem [{\citenamefont {de~Levie}(1967)}]{levie1967electrochemical}%
  \BibitemOpen
  \bibfield  {author} {\bibinfo {author} {\bibfnamefont {R.}~\bibnamefont
  {de~Levie}},\ }in\ \href@noop {} {\emph {\bibinfo {booktitle} {Advances in
  electrochemistry and electrochemical engineering}}},\ Vol.~\bibinfo {volume}
  {6}\ (\bibinfo  {publisher} {Wiley-Interscience New York},\ \bibinfo {year}
  {1967})\ pp.\ \bibinfo {pages} {329--397}\BibitemShut {NoStop}%
\bibitem [{\citenamefont {Janssen}(2021)}]{janssen2021transmission}%
  \BibitemOpen
  \bibfield  {author} {\bibinfo {author} {\bibfnamefont {M.}~\bibnamefont
  {Janssen}},\ }\href {https://doi.org/10.1103/PhysRevLett.126.136002}
  {\bibfield  {journal} {\bibinfo  {journal} {Phys. Rev. Lett.}\ }\textbf
  {\bibinfo {volume} {126}},\ \bibinfo {pages} {136002} (\bibinfo {year}
  {2021})}\BibitemShut {NoStop}%
\bibitem [{\citenamefont {Posey}\ and\ \citenamefont
  {Morozumi}(1966)}]{posey1966theory}%
  \BibitemOpen
  \bibfield  {author} {\bibinfo {author} {\bibfnamefont {F.}~\bibnamefont
  {Posey}}\ and\ \bibinfo {author} {\bibfnamefont {T.}~\bibnamefont
  {Morozumi}},\ }\href {https://doi.org/10.1149/1.2423897} {\bibfield
  {journal} {\bibinfo  {journal} {J. Electrochem. Soc.}\ }\textbf {\bibinfo
  {volume} {113}},\ \bibinfo {pages} {176} (\bibinfo {year}
  {1966})}\BibitemShut {NoStop}%
\bibitem [{Note1()}]{Note1}%
  \BibitemOpen
  \bibinfo {note} {Personal communication with Jie Yang and Cheng
  Lian.}\BibitemShut {Stop}%
\bibitem [{\citenamefont {Alizadeh}\ and\ \citenamefont
  {Mani}(2017)}]{alizadeh2017multiscale}%
  \BibitemOpen
  \bibfield  {author} {\bibinfo {author} {\bibfnamefont {S.}~\bibnamefont
  {Alizadeh}}\ and\ \bibinfo {author} {\bibfnamefont {A.}~\bibnamefont
  {Mani}},\ }\href {https://doi.org/10.1021/acs.langmuir.6b03816} {\bibfield
  {journal} {\bibinfo  {journal} {Langmuir}\ }\textbf {\bibinfo {volume}
  {33}},\ \bibinfo {pages} {6205} (\bibinfo {year} {2017})}\BibitemShut
  {NoStop}%
\bibitem [{\citenamefont {Arrowsmith}\ and\ \citenamefont
  {Place}(1992)}]{arrowsmith1992dynamical}%
  \BibitemOpen
  \bibfield  {author} {\bibinfo {author} {\bibfnamefont {D.}~\bibnamefont
  {Arrowsmith}}\ and\ \bibinfo {author} {\bibfnamefont {C.~M.}\ \bibnamefont
  {Place}},\ }\href@noop {} {\emph {\bibinfo {title} {Dynamical systems:
  differential equations, maps, and chaotic behaviour}}},\ Vol.~\bibinfo
  {volume} {5}\ (\bibinfo  {publisher} {CRC Press},\ \bibinfo {year}
  {1992})\BibitemShut {NoStop}%
\bibitem [{\citenamefont {Lu}(1991)}]{lu1990grobman}%
  \BibitemOpen
  \bibfield  {author} {\bibinfo {author} {\bibfnamefont {K.}~\bibnamefont
  {Lu}},\ }\href {https://doi.org/https://doi.org/10.1016/0022-0396(91)90017-4}
  {\bibfield  {journal} {\bibinfo  {journal} {J. Differ. Equ.}\ }\textbf
  {\bibinfo {volume} {93}},\ \bibinfo {pages} {364} (\bibinfo {year}
  {1991})}\BibitemShut {NoStop}%
\bibitem [{\citenamefont {te~Vrugt}\ \emph {et~al.}(2020)\citenamefont
  {te~Vrugt}, \citenamefont {L{\"o}wen},\ and\ \citenamefont
  {Wittkowski}}]{te2020classical}%
  \BibitemOpen
  \bibfield  {author} {\bibinfo {author} {\bibfnamefont {M.}~\bibnamefont
  {te~Vrugt}}, \bibinfo {author} {\bibfnamefont {H.}~\bibnamefont
  {L{\"o}wen}},\ and\ \bibinfo {author} {\bibfnamefont {R.}~\bibnamefont
  {Wittkowski}},\ }\href {https://doi.org/10.1080/00018732.2020.1854965}
  {\bibfield  {journal} {\bibinfo  {journal} {Adv. Phys.}\ }\textbf {\bibinfo
  {volume} {69}},\ \bibinfo {pages} {121} (\bibinfo {year} {2020})}\BibitemShut
  {NoStop}%
\bibitem [{\citenamefont {Whitaker}(2013)}]{whitaker2013fundamental}%
  \BibitemOpen
  \bibfield  {author} {\bibinfo {author} {\bibfnamefont {S.}~\bibnamefont
  {Whitaker}},\ }\href@noop {} {\emph {\bibinfo {title} {Fundamental principles
  of heat transfer}}}\ (\bibinfo  {publisher} {Elsevier},\ \bibinfo {year}
  {2013})\ \bibinfo {note} {p. 162}\BibitemShut {NoStop}%
\bibitem [{\citenamefont {Corkill}\ and\ \citenamefont
  {Rosenhead}(1939)}]{corkill1939distribution}%
  \BibitemOpen
  \bibfield  {author} {\bibinfo {author} {\bibfnamefont {A.}~\bibnamefont
  {Corkill}}\ and\ \bibinfo {author} {\bibfnamefont {L.}~\bibnamefont
  {Rosenhead}},\ }\href {https://doi.org/10.1098/rspa.1939.0111} {\bibfield
  {journal} {\bibinfo  {journal} {Proc. R. Soc. A}\ }\textbf {\bibinfo {volume}
  {172}},\ \bibinfo {pages} {410} (\bibinfo {year} {1939})}\BibitemShut
  {NoStop}%
\bibitem [{\citenamefont {Dukhin}(1993)}]{dukhin1993non}%
  \BibitemOpen
  \bibfield  {author} {\bibinfo {author} {\bibfnamefont {S.}~\bibnamefont
  {Dukhin}},\ }\href {https://doi.org/10.1016/0001-8686(93)80021-3} {\bibfield
  {journal} {\bibinfo  {journal} {Adv. Colloid Interface Sci.}\ }\textbf
  {\bibinfo {volume} {44}},\ \bibinfo {pages} {1} (\bibinfo {year}
  {1993})}\BibitemShut {NoStop}%
\bibitem [{\citenamefont {Bazant}\ \emph {et~al.}(2004)\citenamefont {Bazant},
  \citenamefont {Thornton},\ and\ \citenamefont {Ajdari}}]{bazant2004diffuse}%
  \BibitemOpen
  \bibfield  {author} {\bibinfo {author} {\bibfnamefont {M.~Z.}\ \bibnamefont
  {Bazant}}, \bibinfo {author} {\bibfnamefont {K.}~\bibnamefont {Thornton}},\
  and\ \bibinfo {author} {\bibfnamefont {A.}~\bibnamefont {Ajdari}},\ }\href
  {https://doi.org/10.1103/PhysRevE.70.021506} {\bibfield  {journal} {\bibinfo
  {journal} {Phys. Rev. E}\ }\textbf {\bibinfo {volume} {70}},\ \bibinfo
  {pages} {021506} (\bibinfo {year} {2004})}\BibitemShut {NoStop}%
\bibitem [{\citenamefont {Kilic}\ \emph
  {et~al.}(2007{\natexlab{a}})\citenamefont {Kilic}, \citenamefont {Bazant},\
  and\ \citenamefont {Ajdari}}]{kilic2007steric}%
  \BibitemOpen
  \bibfield  {author} {\bibinfo {author} {\bibfnamefont {M.~S.}\ \bibnamefont
  {Kilic}}, \bibinfo {author} {\bibfnamefont {M.~Z.}\ \bibnamefont {Bazant}},\
  and\ \bibinfo {author} {\bibfnamefont {A.}~\bibnamefont {Ajdari}},\ }\href
  {https://doi.org/10.1103/PhysRevE.75.021502} {\bibfield  {journal} {\bibinfo
  {journal} {Physical Rev. E}\ }\textbf {\bibinfo {volume} {75}},\ \bibinfo
  {pages} {021502} (\bibinfo {year} {2007}{\natexlab{a}})}\BibitemShut
  {NoStop}%
\bibitem [{Note2()}]{Note2}%
  \BibitemOpen
  \bibinfo {note} {We deduced $H/\lambda _D=20$, which was not reported in
  Ref.~\cite {PRL2014MTL}, from Fig.~4 therein.}\BibitemShut {Stop}%
\bibitem [{Note3()}]{Note3}%
  \BibitemOpen
  \bibinfo {note} {The PNP equations are probably not accurate for the larger
  $\Phi $ values in \protect \cref {fig:charging_dynamics}(b); we consider
  $\Phi <\protect \leavevmode {\protect \color {red}8}$ to compare to
  Ref.~\cite {PRL2014MTL}.}\BibitemShut {Stop}%
\bibitem [{Note4()}]{Note4}%
  \BibitemOpen
  \bibinfo {note} {Reference~\cite {PRL2014MTL} scales $t_\protect \text {ch}$
  by $\tau _{TL}$. The $t_\protect \text {ch}/\tau _{TL}$ data in their
  Fig.~5(d) should approach 1.78 for small applied potential, but it approaches
  1, instead.}\BibitemShut {Stop}%
\bibitem [{\citenamefont {Janssen}\ and\ \citenamefont
  {Bier}(2018)}]{janssen2018transient}%
  \BibitemOpen
  \bibfield  {author} {\bibinfo {author} {\bibfnamefont {M.}~\bibnamefont
  {Janssen}}\ and\ \bibinfo {author} {\bibfnamefont {M.}~\bibnamefont {Bier}},\
  }\href {https://doi.org/10.1103/PhysRevE.97.052616} {\bibfield  {journal}
  {\bibinfo  {journal} {Phys. Rev. E}\ }\textbf {\bibinfo {volume} {97}},\
  \bibinfo {pages} {052616} (\bibinfo {year} {2018})}\BibitemShut {NoStop}%
\bibitem [{\citenamefont {Holmes}(2012)}]{holmes2012introduction}%
  \BibitemOpen
  \bibfield  {author} {\bibinfo {author} {\bibfnamefont {M.~H.}\ \bibnamefont
  {Holmes}},\ }\href@noop {} {\emph {\bibinfo {title} {Introduction to
  perturbation methods}}},\ Vol.~\bibinfo {volume} {20}\ (\bibinfo  {publisher}
  {Springer Science \& Business Media},\ \bibinfo {year} {2012})\BibitemShut
  {NoStop}%
\bibitem [{\citenamefont {Newman}\ and\ \citenamefont
  {Tiedemann}(1975)}]{newman1975porous}%
  \BibitemOpen
  \bibfield  {author} {\bibinfo {author} {\bibfnamefont {J.}~\bibnamefont
  {Newman}}\ and\ \bibinfo {author} {\bibfnamefont {W.}~\bibnamefont
  {Tiedemann}},\ }\href {https://doi.org/10.1002/aic.690210103} {\bibfield
  {journal} {\bibinfo  {journal} {AIChE Journal}\ }\textbf {\bibinfo {volume}
  {21}},\ \bibinfo {pages} {25} (\bibinfo {year} {1975})}\BibitemShut {NoStop}%
\bibitem [{\citenamefont {Biesheuvel}\ \emph {et~al.}(2011)\citenamefont
  {Biesheuvel}, \citenamefont {Fu},\ and\ \citenamefont
  {Bazant}}]{biesheuvel2011diffuse}%
  \BibitemOpen
  \bibfield  {author} {\bibinfo {author} {\bibfnamefont {P.~M.}\ \bibnamefont
  {Biesheuvel}}, \bibinfo {author} {\bibfnamefont {Y.}~\bibnamefont {Fu}},\
  and\ \bibinfo {author} {\bibfnamefont {M.~Z.}\ \bibnamefont {Bazant}},\
  }\href {https://doi.org/10.1103/PhysRevE.83.061507} {\bibfield  {journal}
  {\bibinfo  {journal} {Phys. Rev. E}\ }\textbf {\bibinfo {volume} {83}},\
  \bibinfo {pages} {061507} (\bibinfo {year} {2011})}\BibitemShut {NoStop}%
\bibitem [{\citenamefont {de~Souza}\ and\ \citenamefont
  {Bazant}(2020)}]{souza2020continuum}%
  \BibitemOpen
  \bibfield  {author} {\bibinfo {author} {\bibfnamefont {J.~P.}\ \bibnamefont
  {de~Souza}}\ and\ \bibinfo {author} {\bibfnamefont {M.~Z.}\ \bibnamefont
  {Bazant}},\ }\href {https://doi.org/10.1021/acs.jpcc.0c01261} {\bibfield
  {journal} {\bibinfo  {journal} {J. Phys. Chem. C}\ }\textbf {\bibinfo
  {volume} {124}},\ \bibinfo {pages} {11414} (\bibinfo {year}
  {2020})}\BibitemShut {NoStop}%
\bibitem [{\citenamefont {Kilic}\ \emph
  {et~al.}(2007{\natexlab{b}})\citenamefont {Kilic}, \citenamefont {Bazant},\
  and\ \citenamefont {Ajdari}}]{kilic2007stericII}%
  \BibitemOpen
  \bibfield  {author} {\bibinfo {author} {\bibfnamefont {M.~S.}\ \bibnamefont
  {Kilic}}, \bibinfo {author} {\bibfnamefont {M.~Z.}\ \bibnamefont {Bazant}},\
  and\ \bibinfo {author} {\bibfnamefont {A.}~\bibnamefont {Ajdari}},\ }\href
  {https://doi.org/10.1103/PhysRevE.75.021503} {\bibfield  {journal} {\bibinfo
  {journal} {Phys. Rev. E}\ }\textbf {\bibinfo {volume} {75}},\ \bibinfo
  {pages} {021503} (\bibinfo {year} {2007}{\natexlab{b}})}\BibitemShut
  {NoStop}%
\end{thebibliography}
\end{document}